\documentclass[twocolumn]{aastex63}

\received{}
\revised{}
\accepted{}
\submitjournal{ApJ}

\shorttitle{SL analysis of RMJ1212}
\shortauthors{Zitrin et al.}
\begin{document}

\title{A Strong-Lensing Model for the WMDF JWST/GTO Very Rich Cluster Abell 1489}

\correspondingauthor{Adi Zitrin}
\email{adizitrin@gmail.com}

\author[0000-0002-0350-4488]{Adi Zitrin}
\affiliation{Physics Department,
Ben-Gurion University of the Negev, P.O. Box 653,
Be'er-Sheva 84105, Israel}

\author[0000-0003-3108-9039]{Ana Acebron}
\affiliation{Physics Department,
Ben-Gurion University of the Negev, P.O. Box 653,
Be'er-Sheva 84105, Israel}

\author[0000-0001-7410-7669]{Dan Coe}
\affiliation{Space Telescope Science Institute, 3700 San Martin Dr., Baltimore, MD 21218, USA}

\author[0000-0003-3142-997X]{Patrick L. Kelly}
\affiliation{School of Physics and Astronomy, University of Minnesota, 116 Church Street SE, Minneapolis, MN 55455, USA}

\author[0000-0002-6610-2048]{Anton M. Koekemoer}
\affiliation{Space Telescope Science Institute, 3700 San Martin Dr., Baltimore, MD 21218, USA}

\author[0000-0001-6342-9662]{Mario Nonino}
\affiliation{INAF - Osservatorio Astronomico di Trieste, Via Tiepolo 11, I-34131 Trieste, Italy}

\author[0000-0001-8156-6281]{Rogier A. Windhorst}
\affiliation{School of Earth and Space Exploration, Arizona State University, Tempe, AZ 85287-1404, USA}

\author[0000-0003-1625-8009]{Brenda Frye}
\affiliation{Department of Astronomy, Steward Observatory, University of Arizona, 933 North Cherry Avenue, Tucson, AZ, 85721, USA}

\author[0000-0002-2282-8795]{Massimo Pascale}
\affiliation{Department of Astronomy, University of California, Berkeley, CA 94720-3411, USA}

\author[0000-0002- 8785-8979]{Tom Broadhurst}
\affiliation{Department of Theoretical Physics, University of the Basque Country UPV/EHU, E-48080 Bilbao, Spain}
\affiliation{Donostia International Physics Center (DIPC), 20018 Donostia-San Sebastian (Gipuzkoa), Spain}
\affiliation{Ikerbasque, Basque Foundation for Science, E-48011 Bilbao, Spain}

\author[0000-0003-3329-1337]{Seth H. Cohen} 
\affiliation{School of Earth and Space Exploration, Arizona State University, Tempe, AZ 85287-1404, USA}

\author[0000-0001-9065-3926]{Jose M. Diego}
\affiliation{Instituto de F\'isica de Cantabria (CSIC-UC).  Edificio Juan Jord\'a. Avda Los Castros s/n. 39005 Santander, Spain }

\author[0000-0001-8519-1130]{Steven L. Finkelstein} 
\affiliation{Department of Astronomy, The University of Texas at Austin, Austin, TX 78712}

\author[0000-0003-1268-5230]{Rolf A. Jansen}
\affiliation{School of Earth and Space Exploration, Arizona State University, Tempe, AZ 85287-1404, USA}

\author[0000-0003-2366-8858]{Rebecca L. Larson} 
\affiliation{Department of Astronomy, The University of Texas at Austin, Austin, TX 78712}

\author[0000-0001-7592-7714]{Haojing Yan} 
\affiliation{Department of Physics and Astronomy, University of Missouri, Columbia, MO 65211, USA}

\author[0000-0003-0321-1033]{Mehmet Alpaslan}
\affiliation{Center for Cosmology and Particle Physics, Department of Physics, New York University, New York, NY 10012, USA}

\author[0000-0003-1949-7638]{Christopher J. Conselice} 
\affiliation{School of Physics and Astronomy, The University of Nottingham, University Park, Nottingham, NG7 2RD, UK}

\author[0000-0003-1880-3509]{Alex Griffiths} 
\affiliation{School of Physics and Astronomy, The University of Nottingham, University Park, Nottingham, NG7 2RD, UK}

% \author[0000-0003-1947-687X]{Steve Rodney} 
% \affiliation{Department of Physics and Astronomy, University of South Carolina, 712 Main St., Columbia,
% SC 29208, USA}

\author[0000-0002-7756-4440]{Louis-Gregory Strolger} 
\affiliation{Space Telescope Science Institute, 3700 San Martin Dr., Baltimore, MD 21218, USA}

\author[0000-0001-7956-9758]{J. Stuart B. Wyithe} 
\affiliation{School of Physics, University of Melbourne, Parkville, VIC 3010, Australia}



\begin{abstract}
We present a first strong-lensing model for the galaxy cluster RM J121218.5+273255.1 ($z=0.35$; hereafter RMJ1212; also known as Abell 1489). This cluster is amongst the top 0.1\% richest clusters in the redMaPPer catalog; it is significantly detected in X-ray and through the Sunyaev-Zel'dovich effect in ROSAT and \emph{Planck} data, respectively; and its optical luminosity distribution implies a very large lens, following mass-to-light scaling relations. Based on these properties it was chosen for the Webb Medium Deep Fields (WMDF) JWST/GTO program. In preparation for this program, RMJ1212 was recently imaged with GMOS on Gemini North and in seven optical and near-infrared bands with the \emph{Hubble Space Telescope}. We use these data to map the inner mass distribution of the cluster, uncovering various sets of multiple images. We also search for high-redshift candidates in the data, as well as for transient sources. We find over a dozen high-redshift ($z\gtrsim6$) candidates based on both photometric redshift and the dropout technique. No prominent ($\gtrsim5 \sigma$) transients were found in the data between the two HST visits. Our lensing analysis reveals a relatively large lens with an effective Einstein radius of $\theta_{E}\simeq32 \pm3 \arcsec$ ($z_{s}=2$), in broad agreement with the scaling-relation expectations. RMJ1212 demonstrates that powerful lensing clusters can be selected in a robust and automated way following the light-traces-mass assumption.  
\end{abstract}

%RMJ1212 constitutes another proof-of-concept for a growing ability to approximate the mass-density distribution and the corresponding lensing properties of galaxy clusters with the light-traces-mass assumption, directly -- and automatically -- in large sky surveys.



\section{Introduction}\label{sec:intro}
Strong lensing by galaxy clusters has both enabled studies of the \emph{dark matter} distribution in cluster cores \citep[][for reviews]{Kneib2011review,Bartelmann2010reviewB}, and, a magnified view into the high-redshift galaxies, often not accessible otherwise \citep[e.g.,][]{Franx1997,Frye1998Z4,Ellis2001z5p6,Bradley2008,Coe2012highz,Zheng2012NaturZ,Hashimoto2018Natur.557..392H}.

Deep cluster-lensing campaigns with \emph{Hubble}, such as the Cluster Lensing and Supernova with Hubble (CLASH; \citealt{PostmanCLASHoverview}) and the Reionization Lensing Cluster Survey (RELICS; \citealt{Coe2019RELICS}), have supplied hundreds of high-redshift (z$\gtrsim6$) galaxies in the heart of the reionization era \citep{Bradley2013highz,Salmon2020HighzRelics}. The Hubble Frontier Fields (HFF; \citealt{Lotz2016HFF}) cluster lensing survey targeted six of the most prominent lensing clusters known (with Einstein radii of $\sim$25\arcsec--55\arcsec -- for $z_{s}\simeq2$ sources) in order to maximize the high-redshift science return. The HFF has delivered some of the highest redshift galaxies known to date \citep[e.g.,][]{Laporte2017} and a large sample of the faintest high-redshift galaxies in the reionization era \citep[e.g.,][]{McLeod2015HFF,Zheng2014A2744} reaching as intrinsically faint as $M_{UV}\simeq-16$ at $z\simeq8$ \citep[e.g.,][]{Atek2015HFFhalf,Bouwens2017,Livermore2017,Yue2018HFF,Ishigaki2018HFF}. This is of particular importance because it is believed that faint galaxies are responsible for the reionization of the universe \citep[e.g.,][]{YanWindhorst2004UDF,Robertson15,Finkelstein2016REVIEW,Bouwens2017}. Remarkably, the HFF has not only boosted high-redshift science, but it also led to serendipitous discoveries such as the first resolved multiply imaged supernova \citep{Kelly2015Sci} and first cosmological caustic crossing events of high redshift stars \citep{Kelly2018NatAsCCE,Rodney2018NatAsCCE,Chen2019CCE,Kaurov2019CCE}.

The next leap in these scientific fields is anticipated to take place with the James Webb Space Telescope (JWST). The JWST will enable an extended wavelength coverage and a much deeper view of lensed galaxies \citep[e.g.,][]{Mason2015Observability}, pushing towards intrinsically fainter and higher redshift galaxies -- one of its primary goals. JWST will also reveal many more transient phenomena and caustic crossings in lensing clusters \citep{Windhorst2018ApJSCCE, Venumadhav2017ApJCCE,Oguri2018PhRvDCCE}, enabling important constraints on the composition of dark matter \citep[e.g.,][]{Diego2018ApJCCE,Dai2018CCE}.  As such, to maximize these science cases -- that of high-redshift and reionization in particular -- a number of JWST GTO programs have chosen to concentrate on prominent cluster lenses. 

Most of the prominent cluster lenses known to date were typically chosen following their gas properties: in X-ray -- such as MACS clusters \citep[e.g.,][]{Ebeling2010FinalMACS}, and/or the Sunyaev-Zeldovich (SZ) effect - such as the \emph{Planck} clusters \citep[e.g.,][]{PlanckPSZ22016} imaged in the RELICS program. However, there appear to be many more massive optically selected rich clusters in the sky that are not necessarily bright enough in X-ray or SZ to be included in these samples, but their projected matter distribution is concentrated enough to form a large strong lens \citep[see, e.g.,][]{Wong2012OptLenses}. Such cases (see also \citealt{Umetsu2020WLreview} section 6.2) include the famous Abell 370 HFF cluster with $\theta_{E}\simeq40\arcsec$ \citep{Richard2010A370}; CL0024+1654, with $\theta_{E}\simeq35\arcsec$ \citep{Zitrin2009_cl0024}; or PLCK G165.7+67.0 -- that despite its naming was \emph{not} in fact chosen based on its SZ signal, yet shows an impressive abundance of lensed features \citep{Frye2019}.

The JWST Medium-Deep Fields GTO program (WMDF; PI: Windhorst) has chosen to capitalize on such cases and, in addition to various gas-selected clusters that are well-known prominent lenses, set to observe a few other clusters chosen based on a mix of different probes, including the rich, redMaPPer galaxy cluster RM J121218.5+273255.1 ($z=0.35$; RMJ1212 hereafter; also known as Abell 1489,
RXC J1212.3+2733, or CL1212+2733), as identified in the Sloan Digital Sky Survey \citep[SDSS;][]{York2000SDSS} data. RMJ1212 was provisionally chosen for WMDF mainly as an \emph{optically-selected} strong lens based on the following properties: it is amongst the 0.1\% richest clusters in the redMaPPer catalog \citep[][$\lambda=158.24\pm6.03$ in the public redMaPPer sdss dr8 v6.3 catalog\footnote{\url{http://risa.stanford.edu/redmapper/}}, ranking 33$rd$ out of over $\sim26,000$ clusters]{Rykoff2014Mapper}; it has a high ecliptic latitude ($b\gtrsim30^{\circ}$) minimizing zodiacal near-infrared background light (an important consideration for JWST high-redshift targets); it had a prominent lensing strength, and large Einstein radius of $\sim40\arcsec\pm20\%$ predicted from mass-to-light rescaling of SDSS clusters \citep{Zitrin2012UniversalRE}; its preliminary velocity dispersion estimate from few measured cluster members in SDSS is high, over $\sim900$ km/s; and given that it is significantly detected in X-ray (e.g., 6.13$\sigma$ in ROSAT; or an estimated mass of M$_{500}=10.3\pm2.1\times10^{14}$ M$_{\odot}$, \citealt{Mantz2010}), and SZ (10.09$\sigma$ in \emph{Planck}; an SZ mass proxy of M$_{500}=7.50\pm0.44\times10^{14}$ M$_{\odot}$\footnote{this is only somewhat lower than the cut for the SZ-selected RELICS sample, $8.7\times10^{14}$ M$_{\odot}$.}, \citealt{PlanckPSZ22016}).

\begin{figure*}
\centering
\includegraphics[width=80mm,height=80mm,trim={0cm 0cm 0cm 0cm},clip]{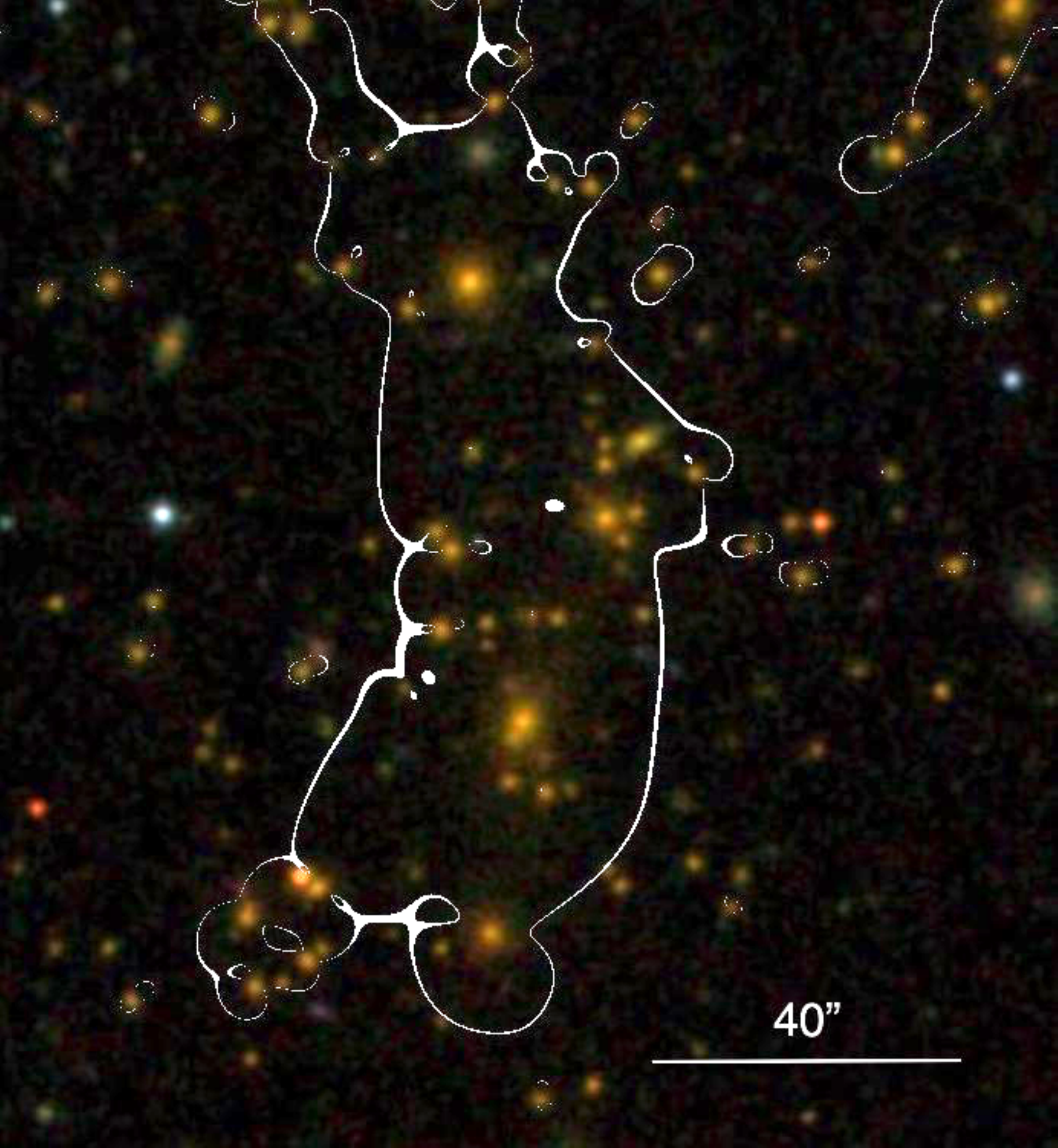}
\includegraphics[width=80mm,height=80mm,trim={0cm 0cm 0cm 0cm},clip]{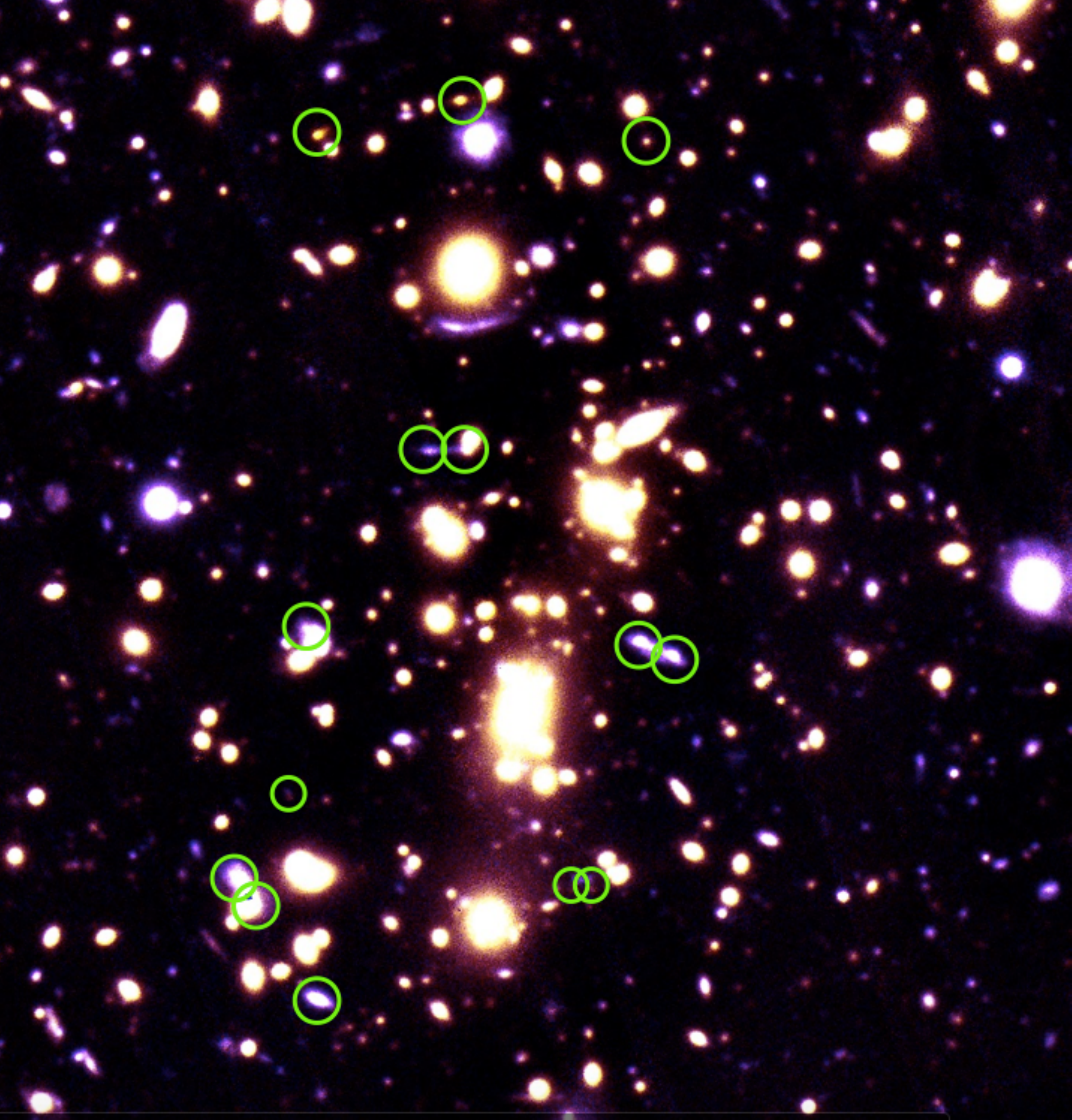}
%\captionsetup{labelformat=empty}
\caption{Preliminary lensing properties of RMJ1212 from ground-based data. \emph{Left:} prediction for the shape and size of the critical area in RMJ1212 based solely on our mass-to-light scaling relation of SDSS selected clusters (\citealt{Zitrin2012UniversalRE}; white lines mark the critical curves for $z_{s}=2$). The lens was predicted to be large, with an Einstein radius of $\sim40\arcsec$. The typical error in this estimate is about 20\%. \emph{Right:} the Gemini/GMOS deep $gr$ observations of RMJ1212. Marked on the image are candidate multiply imaged galaxy families as identified prior to the HST data (the $i$ band observations were not yet available at the time of the candidate multiple image identification and hence were not included). Their location follows nicely the critical curves predicted, suggesting that indeed RMJ1212 is a prominent lens, but HST observations were required to corroborate this, secure these identifications and construct a credible lens model for the cluster, which are our goals here}.% In Fig. \ref{fig:curve}, we show the revised analysis enabled by the \emph{Hubble} data, which is essential for multiple-image identification.}
\label{fig:prelim}
\end{figure*}

\begin{figure*}
 \begin{center}
 %\vspace{0.2cm}
  \includegraphics[width=190mm,height=170mm,trim=4cm 1cm 1cm 1cm,clip]{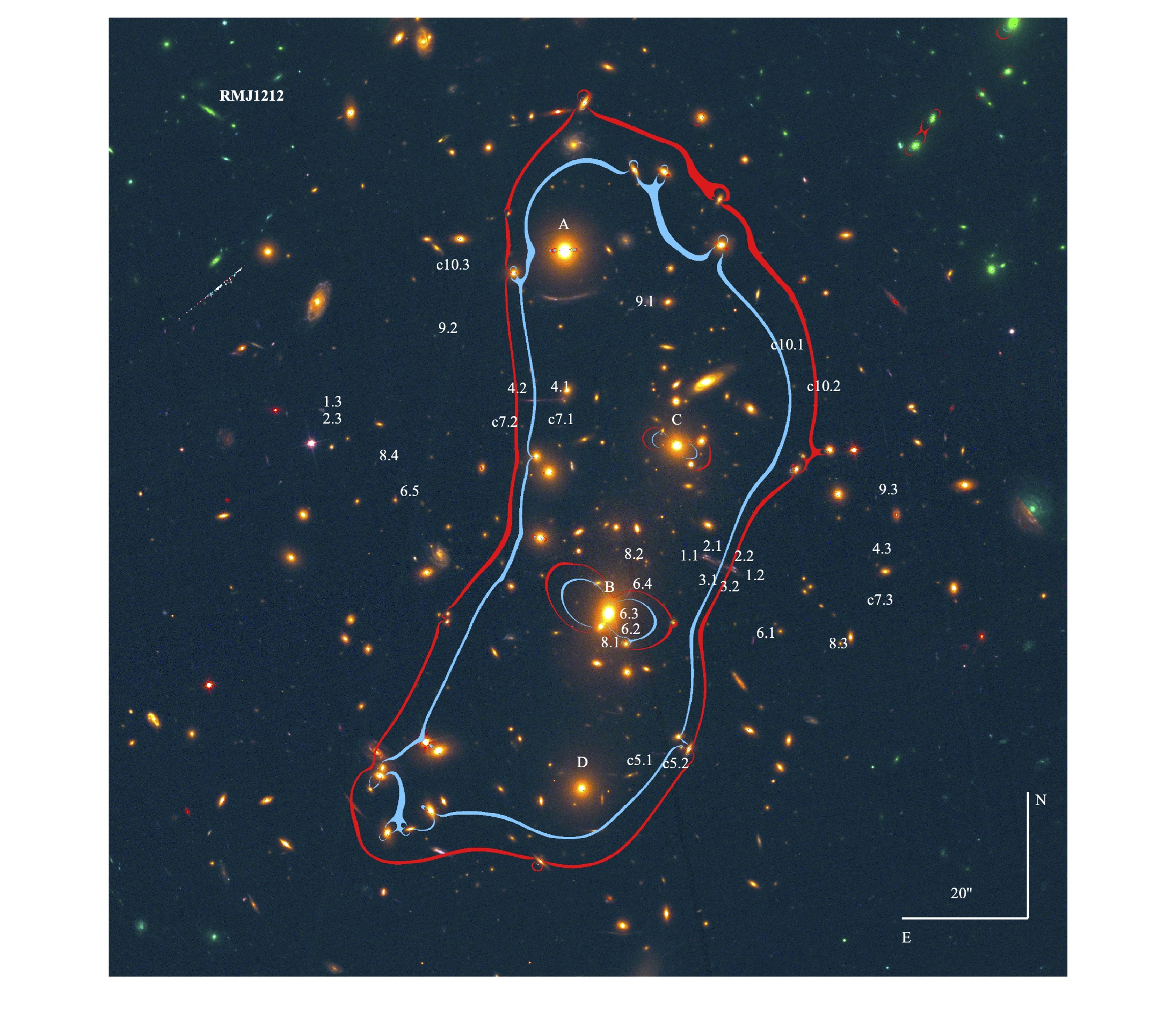}
 \end{center}
\caption{Multiple images and critical curves for RMJ1212. Shown is an RGB color-composite image from the 7-band HST data. Multiple-image sets are marked on the image and the critical curves from our model are overlaid in \emph{light blue} for a source at a redshift similar to that of the main arc (system 1; $z_{phot}\sim2.7$), and in \emph{red} for a source redshift of $z_{s}\simeq10$. Bright cluster members, some of whose parameters were individually optimized in the minimization (see text), are noted with A, B, C and D.}\vspace{0.1cm}
\label{fig:curve}
\end{figure*}

Following these promising properties, we targeted RMJ1212 with GMOS on Gemini-N in imaging mode and detected various sets of potentially multiply imaged galaxies (see Fig. \ref{fig:prelim}). Since space data is typically required to verify the tentative identification of multiple images, RMJ1212 was also recently imaged with the \emph{Hubble Space Telescope} in seven bands (see Fig. \ref{fig:curve} and \S \ref{s:data} for details). Here, we present a first strong-lensing analysis of the cluster in these data. We also search these data for bright, high-redshift dropout galaxies lensed by the cluster and for transient events appearing between the two \emph{HST} visits of the cluster. Updated lens models, including with spectroscopic redshifts for the multiple images\footnote{Spectroscopic observations were already in queue for GMOS/Gemini-North before the telescope shut down on March 2020 due to the COVID-19 pandemic, and were eventually completed on June 2020 after the writing of this manuscript. These will be reduced and incorporated in a future model.}, and extending out to the weak-lensing regime, are planned for future work (Pascale et al., \emph{in preparation}).

% Clusters of galaxies, the most massive bound objects in the universe, are typically found to lens background galaxies that lie behind them, close to the line of sight --> dark matter and high redshift galaxies.
% In search for the best lenses that maximize high-redshift sources......X-ray, clash, relics, frontier fields...
%  Optical selection... 
 
% scaling relation, carrasco...

% The Webb Medium-Deep Fields...

% RMJ1212 and as such was chosen....

% Here, we present a first strong lensing analysis of the cluster in recent imaging data taken with the GMOS on Gemini-North, and the \emph{Hubble Space Telescope}. This paper presents a first working model for the cluster. Updated models, including spectroscopic redshifts for the multiple images\footnote{Spectroscopic observations were already in queue for GMOS/Gemini-North before the telescope shut down on March 2020 due to the COVID-19 pandemic. We anticipate these data will be taken in the next available time frame.}, and extended to the weak lensing regime, are expected in the near future. 

The paper is organized as follows: In \S \ref{s:data} we detail the observations of the cluster and their data reduction. \S \ref{s:code} outlines the Light-Traces-Mass (LTM) lens modeling code, and its implementation to RMJ1212. The modeling results are presented and discussed in \S \ref{s:results}, and concluded in \S \ref{s:summary}. Throughout this work we use a $\Lambda$CDM cosmology with $\Omega_{M}=0.3$, $\Omega_{M}=0.7$, and $H_{0}=70$ km/s/Mpc. Unless otherwise stated, we generally use AB magnitudes \citep{Oke1983ABandStandards}, and errors are $1\sigma$.

\section{Observations and Data Reduction}\label{s:data}
The galaxy cluster RMJ1212 was imaged in queue mode with GMOS on Gemini-N (program ID: GN-2019A-Q-903, PI: Zitrin) on 2019 March 12, in the $g$ ($8\times 600 $s)  and $r$ ($\simeq7.5 \times 600 s$) bands, and on 2019 April 04, in the $i$ ($8 \times 300 s$) band. Data were retrieved from the Gemini archive and reduced using standard procedures with the Gemini \textsc{Iraf} pipeline. Astrometry, based on SDSS,  was obtained using \textsc{Scamp} \citep{Bertin2006SCAMP}. The images, after background subtraction, were coadded using \textsc{Swarp} \citep{Bertin2002TERAPIX}, and zero points for the {\em g,r,i} stacks were obtained via a comparison with SDSS. A color composite image of these data is shown in Figure \ref{fig:prelim}, \emph{right panel}. 

In preparation for the WMDF JWST/GTO program, and in order to construct a detailed lens model, identify high-redshift candidates, and form baseline observations for future transient detection, RMJ1212 was recently observed with \emph{Hubble} for five orbits (program ID: 15959, PI: Zitrin). The cluster was observed for a total of about 1 orbit in each of the F435W (1934s), F606W (1904s), and F814W (1934s) filters with the \emph{Advanced Camera for Surveys (ACS) Wide Field Channel (WFC)}; and for a total of about $1/2$ an orbit in each of the F105W (1212s), F125W (912s), F140W (912s), and F160W (1612s) filters with the \emph{Wide-Field Camera 3 - infrared channel (WFC3/IR)}. One F140W exposure suffered from a guiding problem. To fix this, the readout ($\sim50s$) in which the drift was detected was discarded, and not used in constructing the final images. Observations were divided into two visits, both to relax scheduling constraints, and to allow -- albeit at low probability -- for transient searches, and were carried out on 2020 March 16 and 2020 March 25. 

Mosaic images for all the HST exposures were produced from the calibrated exposures by running the MosaicDrizzle pipeline \citep{Koekemoer2011}, specialized for this proposal and updated to use the latest drizzlepac routines, achieving milliarcsecond-level astrometric alignment across all the different filters for ACS/WFC and WFC3/IR. Two sets of mosaics were produced, at scales of 0$\farcs$03 and 0$\farcs$06 per pixel, with all the pixels aligned to the same astrometric grid, with North up, and registered onto the Gaia DR2 reference frame. The 0$\farcs$03 mosaics are most useful for studying fine morphological details, especially since they sub-sample the native ACS pixel scale, while the 0$\farcs$06 mosaics are more generally useful for producing catalogs across both WFC3/IR and ACS, which we describe in this section.

Throughout we work with several photometric catalogs. First, for the lens model, we need a list of potential cluster members and their photometry. We run \textsc{SExtractor} \citep{BertinArnouts1996Sextractor} version 2.25.0 on the F814W band, and then in dual mode on all other HST images with the F814W as the detection image. Cluster members are identified following the red-sequence in a color-magnitude diagram, where we use the F606W-F814W color versus the F814W magnitude. Members are chosen within $\pm0.15$ mag from this sequence (defined as $(mag_{F606W}-mag_{F814W})=0.05*mag_{F814W}+0.29$), and up to a magnitude of $mag_{F814W}=22.5$. The list of cluster members is then examined visually, and updated as needed (we also include an apparently foreground, $z_{phot}\simeq0.2$ bright galaxy near the cluster center, but with 1/20 weight, i.e., a fraction of the mass implied by its flux). This photometric catalog is not used for any other purpose. 

We also produced a set of catalogs using the RELICS
pipeline \citep{Coe2019RELICS}, which generates combined RGB images from various bands, and runs both SExtractor (version
2.8.6) and the Bayesian Photometric Redshift code \citep[BPZ;][]{Benitez2004,Coe2006}.  Photometry is corrected for Galactic extinction using the IR dust emission maps of \citet{SchlaflyFinkbeiner2011Extinction}. Note that with seven bands spanning a wider wavelength range and more accurately calibrated data than the ground based data, only the HST data are used for fitting photometric redshifts. The RELICS pipeline creates two source catalogs: ``acs-wfc3ir", based on detections in a weighted stack of all HST images (ACS$+$WFC3/IR), optimized to detect most objects; and ''wfc3ir", based on detections in a weighted stack of the WFC3/IR images, and using a finer background grid and more aggressive deblending, optimized to detect smaller high-redshift galaxies. Independently, we also build another alternative catalog based on HST photometry to search for faint high-redshift galaxies. We run SExtractor in dual-mode the seven HST bands, with a similarly weighted stack of all WFC3/IR images as the detection image. We use a local background estimate with $back\_size=16$, $detect\_minarea=8$, $detect\_thresh=1$, $analysis\_thresh=1.5$ and $deblend\_nthresh=16$ to improve source detection. Note that all magnitudes are measured and given hereafter in isophotal apertures, and corrected for galactic extinction.

\begin{figure}
 \begin{center}
 %\vspace{0.2cm}
  \includegraphics[width=90mm,height=80mm,trim=2.5cm 2cm 3cm 2.5cm,clip]{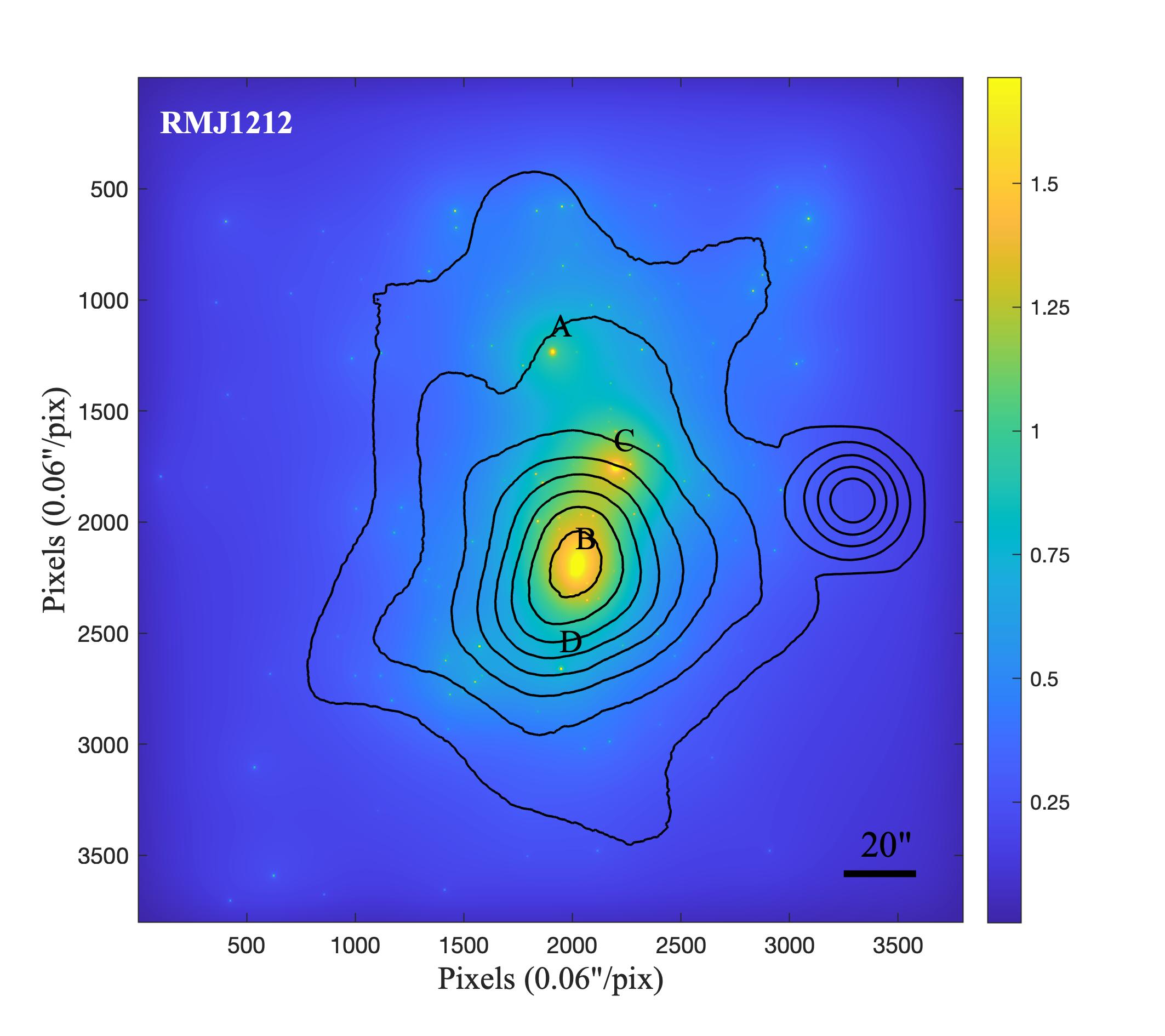}
 \end{center}
 \caption{Projected mass density of RMJ1212. We show
$\kappa$, the surface mass density distribution in units of the critical density for strong lensing, for the source redshift adopted
for the main arc ($z_{s} = 2.7$; systems 1, 2, 3). Overlaid black
contours follow a smoothed X-ray map from Chandra. Capital
letters A, B, C and D mark the four central BCGs as in
Fig. \ref{fig:curve}. From the lensing analysis (and its resulting $\kappa$ map shown) it becomes apparent that the main mass concentration
in centered on galaxy B, in agreement with the X-ray contours,
and despite galaxy A being comparable in optical brightness (and
in fact slightly brighter, see \S \ref{s:results}). The prominent contours on the right hand side correspond to a point source in the X-ray data, coinciding with a prominent spiral galaxy (RA$=$12:12:12.829, DEC$=$+27:33:12.831) and likely suggesting the presence of an Active Galactic Nucleus. X-ray data were interpolated for clarity and should be regarded as qualitative only.}
\label{fig:kappa}
\end{figure}

We also show here a smoothed X-ray map taken with \emph{Chandra} in 2003, January 11 (Obs. ID 5767, PI: Vikhlinin), with an exposure time of 15.0 ks. We use the high resolution ACIS Primary data product, smoothed with a 20-pixel Gaussian. These X-ray data are only used qualitatively to show the X-ray centroid (Fig. \ref{fig:kappa}).

The reduced \emph{HST} images, combined color images and catalogs, are made publicly available online\footnote{\url{https://www.stsci.edu/~koekemoer/zitrin/RMJ1212}}. The lens model now detailed is also available in the same library.

% \textcolor{blue}{Dan is it OK if we make the library public?}.
%http://www.stsci.edu/~dcoe/zitrin/RMJ1212/

%Zero points for the photometry for the WFC3/IR bands are taken from the online instrument documentation\footnote{http://www.stsci.edu/hst/instrumentation/wfc3/data-analysis/photometric-calibration/ir-photometric-calibration}, and time-dependent zero points for the ACS observations were calculated using the online ACS Zeropoints Calculator\footonote{https://acszeropoints.stsci.edu/}.

\section{SL modeling of RMJ1212}\label{s:code}
We use here the Light-Traces-Mass (LTM) lens modeling code of \citet[][and references therein]{Zitrin2009_cl0024,Zitrin2014CLASH25}, which is especially useful for the analysis of new lenses as it is inherently capable of guiding the detection of multiple-image sets \citep{Carrasco2020}. More complete details of the formalism can be found in the above references, and we give here only a broad outline.

The deflection field is modeled as a sum of three components. The first component maps the projected mass density distribution of the cluster galaxies, modeled each with a simple power-law ($q$) surface-mass density, scaled with the galaxy's luminosity and normalized to a desired lensing distance, or redshift, by some factor ($K$). The second component is the dark matter map which is obtained by smoothing the galaxy map with a Gaussian kernel of width $S$. The two components are then added with a relative weight $k_{gal}$, reflecting the ratio of luminous to dark matter. The third component, contributing only to the deflection field, but not to the mass density, is an external shear of strength $\gamma_{ex}$ and position angle $\phi_{ex}$, which allows for greater effective elongation of the critical curves and accounts for the contribution of larger-scale structure. The model thus comprises six main parameters: $q,S,K,k_{gal},\gamma_{ex}$ and $\phi_{ex}$.

We often introduce ellipticites and position angles as well as central cores for a few key cluster members, such as the brightest (and thus most massive) galaxies. These can either be set as fixed, or be minimized as well (adding, correspondingly, to the number of free parameters). In addition, it is often beneficial to leave the relative weight (i.e., the relative mass-to-light ratio) of some key galaxies free as well. Similarly, the lensing distance (i.e., essentially, redshift) of systems with poorly constrained redshifts can also be left to be freely optimized.

The optimization of the model is carried out by minimizing, using a $\chi^2$ function, the distance between multiple images and their positions predicted by the model. This is done via a Monte-Carlo Markov Chain with a Metropolis-Hastings algorithm \citep[e.g.,][]{Hastings1970MCMC}. We also include some annealing in the procedure, and the chain typically runs for several thousand steps after the burn-in stage. Errors are calculated from the same Markov chain.

For modeling RMJ1212 we start with all galaxies fixed, and construct a simple model minimized by two sets of obvious multiple images: images 1.1+1.2, and images 4.1+4.2, fixing them to their best-fit photometric redshifts (Table \ref{multTable}). With this preliminary model we iteratively predict the location of counter images and find additional systems by sending notable arclets across the field to the source plane and back, probing a range of redshifts. In total, we find 31 likely multiple images and candidates of about 8-10 background sources (systems 1, 2 and 3 may correspond to a single background object). With these, we construct the final model presented here. To anchor the model we fix the redshift of all systems to roughly their average, or typical, best BPZ value, listed in Table \ref{multTable} as well (although some minor differences may exist due to updates to the catalog; these only weakly affect the derived $D_{ls}/D_{s}$ ratio). We leave the weight of the four central brightest cluster galaxies (BCGs A,B,C,D in Fig. \ref{fig:curve}) free, and allow for ellipticity for the brightest two (A and B). For the southern BCG around which system 6 appears, BCG ``B", we allow the ellipticity to vary around its values measured from SExtractor. No cores are incorporated for cluster galaxies, aside from this southern one, for which we allow a small core. In total, the model includes 13 free parameters, optimized using uniform flat priors. Image position uncertainties were adopted to be 0.5\arcsec, whereas for systems 1, 3, and 6, we adopted 0.1\arcsec (to give them more weight in the modeling). %The resulting model is described in \S \ref{s:results}. 

\section{Results and Discussion}\label{s:results}

\subsection{Lens modeling}
\begin{figure*}
 \begin{center}
 %\vspace{0.2cm}
  \includegraphics[width=0.3\textwidth,trim=0cm 0cm 0cm 1.1cm,clip]{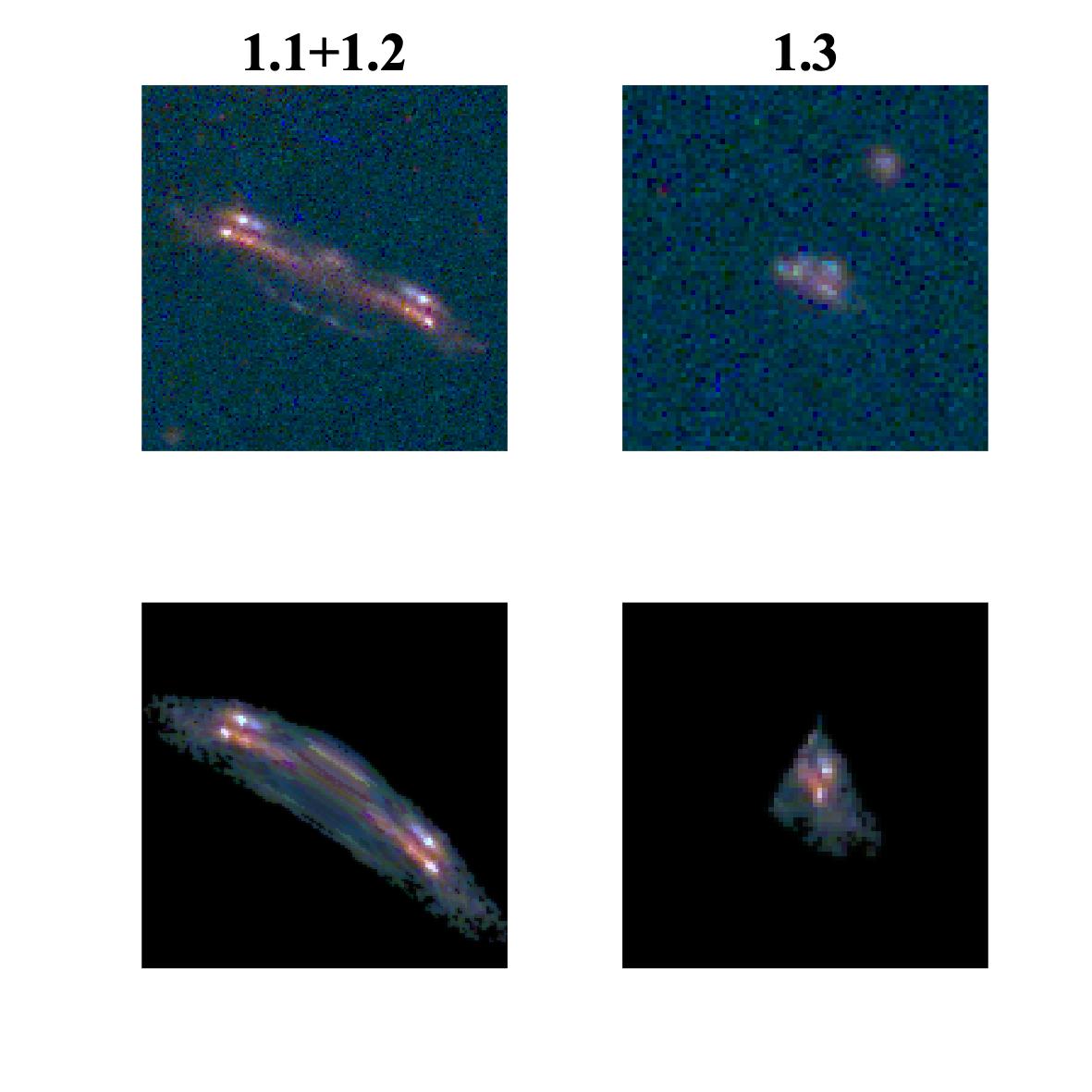}
  \includegraphics[width=0.3\textwidth,trim=0cm 0cm 0cm 1.1cm,clip]{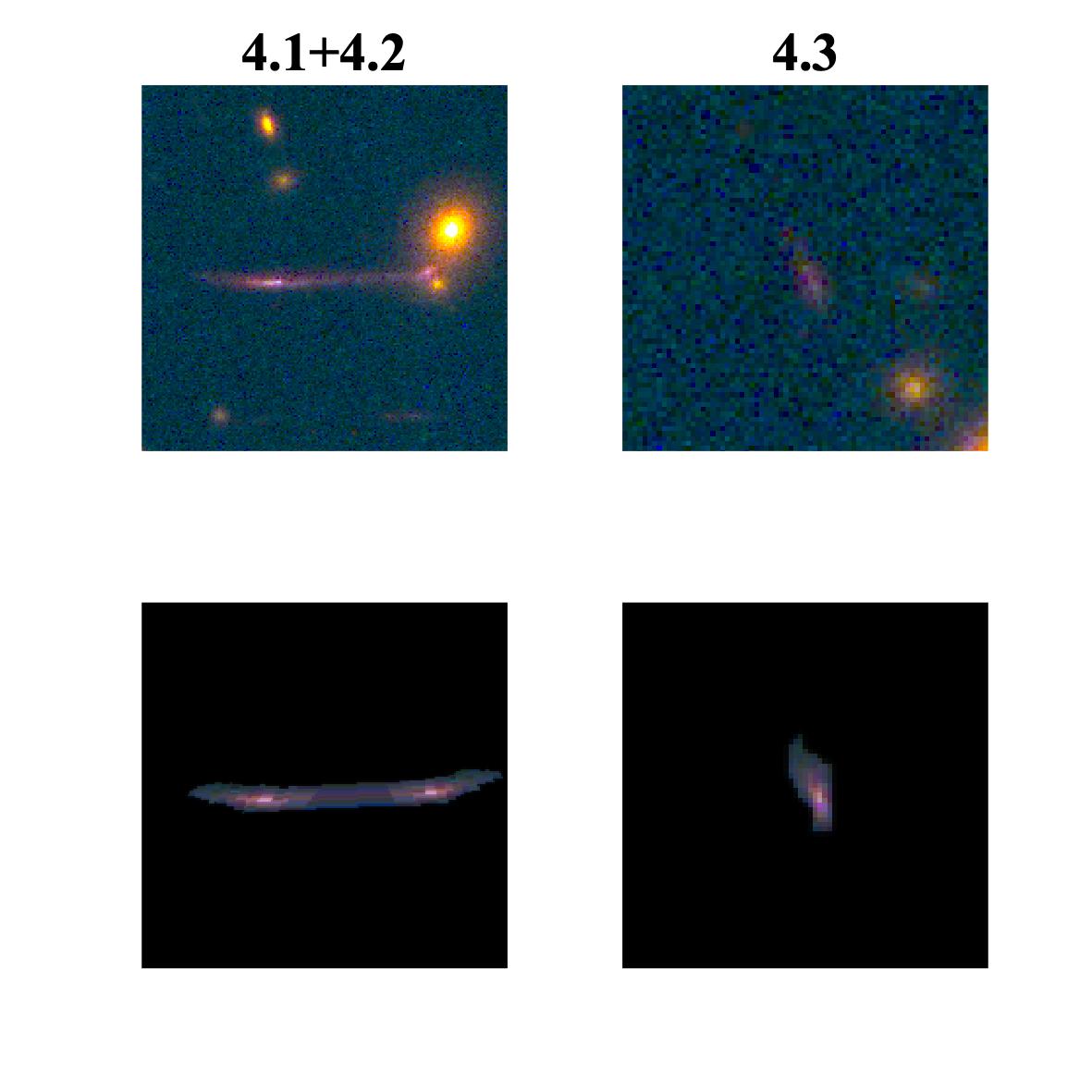}\\
  \includegraphics[width=0.8\textwidth,trim=8cm 0cm 8cm 0.5cm,clip]{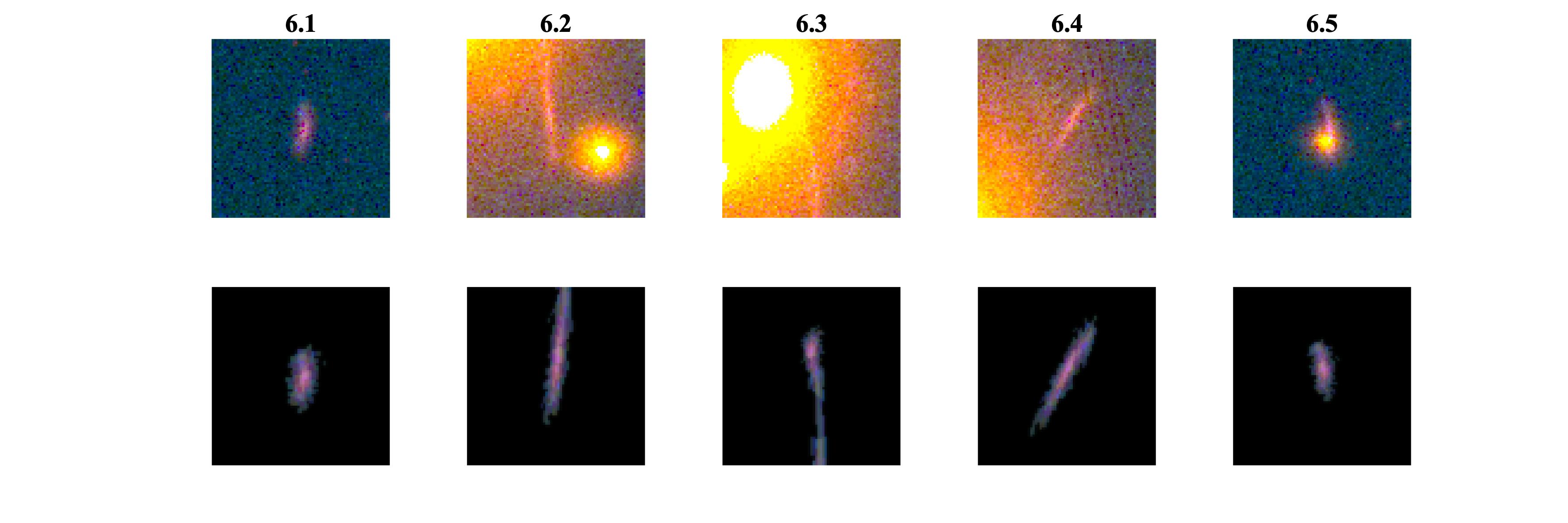}\\
  \includegraphics[width=0.8\textwidth,trim=8cm 0cm 8cm 0.5cm,clip]{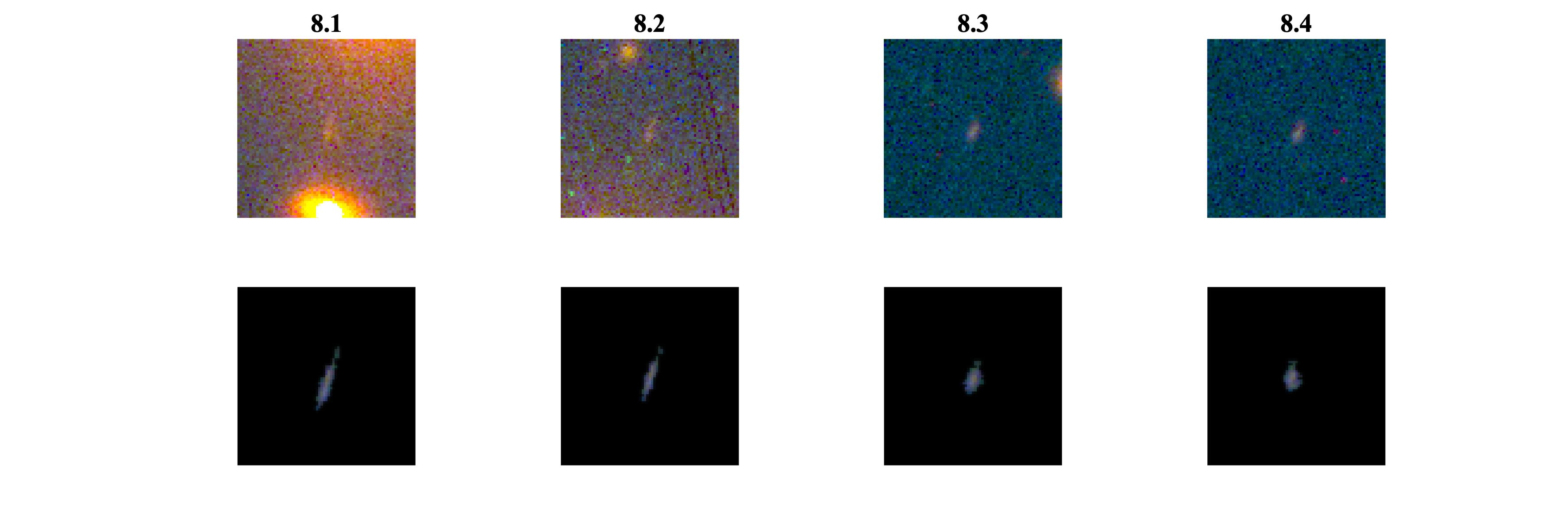}\\
  \includegraphics[width=118mm,trim=0cm 5cm 0.cm 0.35cm,clip]{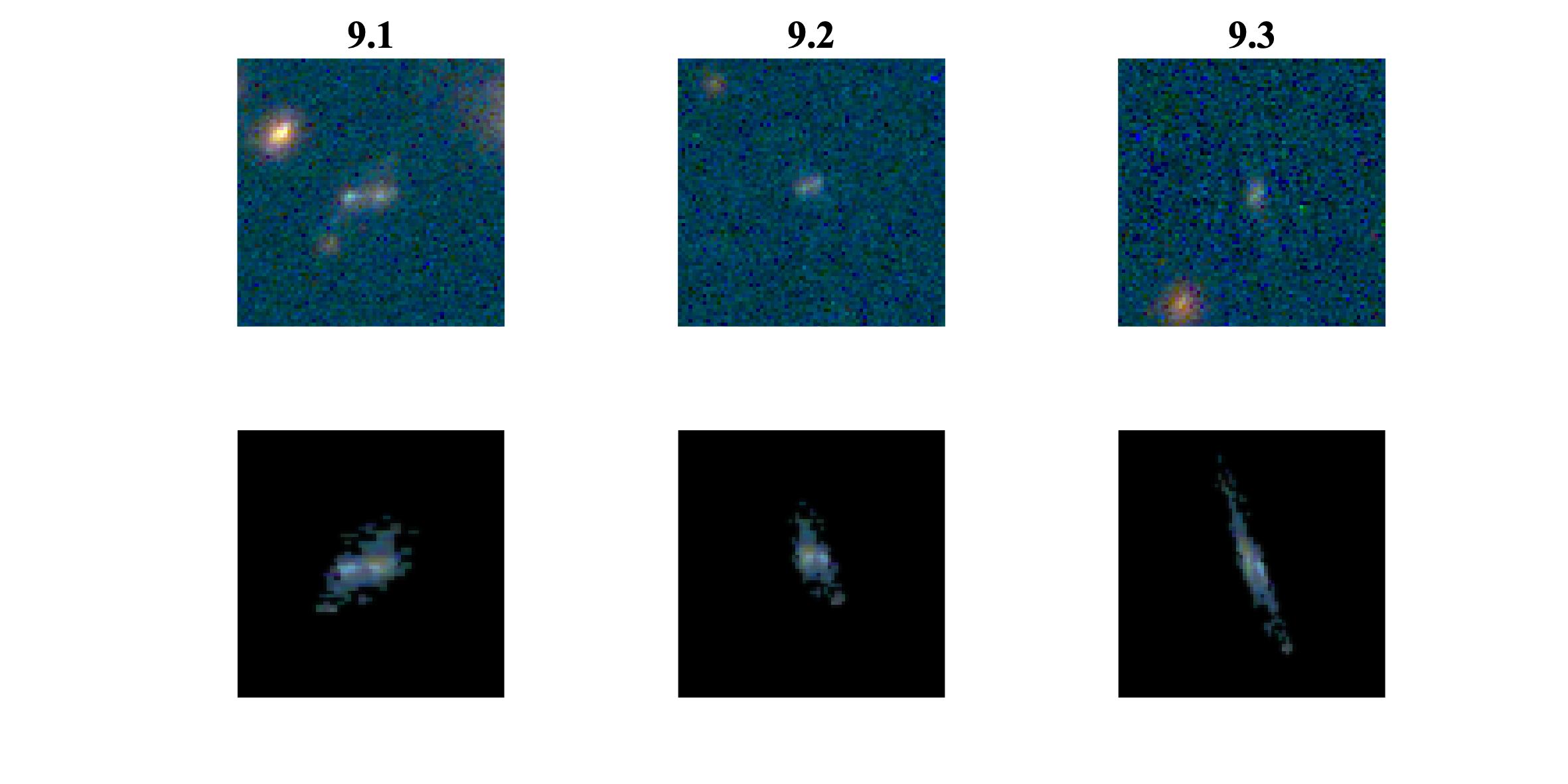}
 \end{center}
\caption{Reproduction of multiple images by our model. We show the reproduction of system 1 (and 2 and 3), by lensing the left side of the arc, image 1.1+2.1+3.1, to the source-plane and back through the lens; system 4, by lensing image 4.1 to the source-plane and back; system 6, by lensing image 6.1 to the source-plane and back; system 8, by lensing image 8.3 to the source-plane and back; and system 9, by lensing image 9.1 to the source-plane and back. For each system the upper row shows the images in the data and the bottom row the reproduction by the model. Although some minor differences exist, the prediction of the model evidently reproduces the observed images well, strengthening their identification.}\vspace{0.1cm}
\label{fig:Reporudiction}
\end{figure*}

The resulting best-fit model has an \emph{rms} of 1\arcsec.44  in reproducing the position of multiple images. Such values are typical for LTM models of moderately complex, large lenses. The reproduction of multiple images is excellent, and a few examples are given in Fig. \ref{fig:Reporudiction}.

The projected mass density map of the best-fit model is shown in Fig. \ref{fig:kappa}. An interesting point to note is that while galaxy A in Fig. \ref{fig:curve} is slightly brighter ($mag_{F814W}=17.37$ AB) than galaxies B ($mag_{F814W}=17.68$ AB) and C ($mag_{F814W}=17.93$ AB), the latter two seem to represent locations of much greater concentrations of mass, with galaxy B being apparently in the center of the potential well of the cluster. This finding agrees very well also with the X-ray signal, which is concentrated on galaxy B (Fig. \ref{fig:kappa}, \emph{black contours}).

The corresponding critical curves of the best-fit model are overlaid on an image of the cluster in Fig. \ref{fig:curve}, where multiple images are marked as well. The effective Einstein radius we find is relatively large, $31.6\pm3.2\arcsec$ for a source at $z_{s}=2$, where the effective Einstein radius is defined as the radius of the area enclosed within the critical curves if it were a circle. The mass in that critical area is $1.53\pm0.21\times10^{14}$ M$_{\odot}$. Errors on the Einstein radius and mass are nominal, systematic values, reflecting typical errors seen between different models; the statistical uncertainties are somewhat smaller. For the assigned redshift of systems 1-3, $z_{s}=2.7$, the light-blue curves shown in Fig. \ref{fig:curve} have an effective $\theta_{E}\simeq34\arcsec$, and for a source at $z_{s}=10$ (red curves therein), they reach $\simeq39\arcsec$. These estimates will be revised once multiple image redshifts become available, and perhaps when more multiple images, especially around the northern end, are identified. 

\begin{deluxetable*}{lccccc}
\tablecaption{Multiple Image Systems}
\label{multTable}
\tablecolumns{6}
%\tablewidth{0.85\linewidth}
\tablehead{
\colhead{ID
} &
\colhead{R.A
} &
\colhead{DEC.
} &
\colhead{$z_{phot}$ [95\% C.I.]
} &
\colhead{$z_{model}$
} &
\colhead{Comment}\\  
 &J2000.0&J2000.0&&  
}
\startdata
1.1 & 12:12:17.315 & +27:33:04.24 & 2.743 [2.642 -- 2.877] & 2.70 & \\
1.2 & 12:12:16.983 & +27:33:02.28 & 2.746 [2.610 -- 2.776] & ''&\\
1.3 & 12:12:21.881 & +27:33:27.42 &  2.885 [2.744 -- 3.105] & ''&\\
\hline
2.1 & 12:12:17.346 & +27:33:03.92  & 2.675 [2.554 -- 2.739] & ''&\\
2.2 & 12:12:16.971 & +27:33:01.72  & 2.600 [2.518 -- 2.719] & ''&\\ 
2.3 & 12:12:21.881 & +27:33:27.14  & 2.754 [2.588 -- 2.978]  & ''&\\ 
\hline
3.1 & 12:12:17.217& +27:33:02.13  & 2.356  [2.097 -- 2.546] & ''&\\
3.2 & 12:12:17.148& +27:33:01.75  & 2.612  [2.446 -- 2.739]  & ''&\\
\hline
4.1 & 12:12:19.028& +27:33:29.16 & 0.529  [0.151 -- 0.706]  & 1.69&\\ 
4.2 & 12:12:19.385& +27:33:28.81  & 1.698 [1.544 -- 1.795] & ''&\\
4.3 & 12:12:15.376& +27:33:04.16 &  1.617 [1.401 -- 1.827]  & ''& \\
\hline
c5.1 & 12:12:17.948& +27:32:33.02 & 1.358 [1.168 -- 1.417]  & ---&\\ 
c5.2 & 12:12:17.810& +27:32:32.92 & 1.201 [1.143 -- 1.352]  & ---&\\
\hline
6.1 & 12:12:16.757& +27:32:50.82 & 1.455 [1.316 -- 1.554]  & 1.34&\\
6.2 & 12:12:18.374& +27:32:51.38 &--- & ''&\\
6.3 & 12:12:18.393& +27:32:53.85 & ---& ''&\\
6.4 & 12:12:18.232& +27:32:58.52 & --- & ''&\\
6.5 & 12:12:21.006& +27:33:13.25 & --- & ''&\\
\hline
c7.1 & 12:12:19.061& +27:33:24.68 & 2.377 [1.467 -- 2.717] & ---&\\
c7.2 & 12:12:19.732& +27:33:23.64 &2.120 [1.560 -- 3.153]    & ---&\\
c7.3 & 12:12:15.396 & +27:32:57.20&  1.151  [0.933 -- 2.347]   & ---&\\
\hline
8.1 & 12:12:18.615& +27:32:49.24 &  ---  & 2.59&\\
8.2 & 12:12:18.334& +27:33:03.27 & --- &  ''&\\
8.3 & 12:12:15.898& +27:32:49.08 &  2.187 [1.433 -- 2.606] &  ''&\\
8.4 & 12:12:21.256& +27:33:18.93  & 2.304  [2.018 -- 2.556]&  ''&\\
\hline
9.1 & 12:12:18.199& +27:33:43.28  & 3.254 [3.094 -- 3.374] & 3.35&\\
9.2 & 12:12:20.552& +27:33:39.08  & 0.568 [0.132 -- 3.189] & ''&\\
9.3 & 12:12:15.295& +27:33:13.47  & 3.402 [3.224 -- 3.503]& ''&\\
\hline
c10.1 & 12:12:16.586 & +27:33:36.49 & 2.578 [0.543 -- 2.738]&---&\\
c10.2 & 12:12:16.153 & +27:33:29.88 &  0.928 [0.600 -- 2.394]&---&\\
c10.3 & 12:12:20.575 & +27:33:49.15 &  2.615 [0.115 -- 3.056]&--- &\\
\hline
\enddata
\tablecomments{Multiple images and candidates. \emph{Column~1:} ID; \emph{Column~2 \& 3:} Right Ascension and Declination, in J2000.0; \emph{Column~4:} best photometric redshift from BPZ, and its 95\% confidence interval; \emph{Column~5:} the redshift of the system as adopted for the modeling; \emph{Column~6:} comments. ''c" stands for candidate image, whose identification was less secure and it was not used in the minimization.}
\end{deluxetable*}

It has been known that substantial substructure projected near the core in merging clusters, as seen in RMJ1212, boosts the Einstein radius \citep{Torri2004,Redlich2012MergerRE}. Searching for the largest and most efficient lenses is important for a variety of studies. Given the shape of the cosmological mass function and the hierarchical mass build up, massive clusters are rarer, and their numbers have direct implications for structure formation and evolution models, as well as for cosmological models. While a total mass does not guarantee a large lens, overall the Einstein radius should increase with the mass of the cluster (albeit with a large scatter), and predictions can be made for the universal Einstein radius distribution based on an input mass function, cosmology, and assumptions on the shape of the clusters \citep[e.g.][]{OguriBlandford2009}. 
In addition, and especially as we prepare for the next generation space telescope, JWST, a primary goal of which is detecting the first galaxies, we wish to find those lenses that maximize the high-redshift galaxy yield. Merging galaxy clusters, especially those with elongated shapes, are known to have a boosted lensing cross section \citep[e.g.,][]{Meneghetti2003,Zitrin2013M0416,Acebron2019CL0152}, such as the HFF clusters \citep{Lotz2016HFF,Vega-Ferrero2019probHFF}, and thus should be favorable for detecting high-redshift galaxies. On the other hand, some massive clusters, despite being merging with many subclumps and comprising large Einstein radii, are not necessarily the most prolific in terms of high-redshift galaxies \citep[see for example][]{Acebron2019RXC0032}, but it is not yet clear if this is a result of cosmic variance, less available area outside the critical curves to search for high-redshift dropouts (see also \citealt{Oesch2015LF_HFF}), or indication of less a steep faint-end luminosity function than what is needed to counter the magnification bias \citep[e.g.,][]{Broadhurst1995MagBias,Mashian2013}. Ongoing surveys such as the BUFFALO survey \citep{Steinhardt2020BUFFALO} mapping larger areas around the Hubble Frontier Fields clusters, should be helpful in answering this important question with \emph{HST}, in the advent of next generation telescopes. 

The largest gravitational lenses known to date (see the list in Table 1 of \citealt{Acebron2019RXC0032}) have been usually chosen for \emph{HST} observations based on their X-ray (MACS clusters and respective snapshot programs, \citealt{Ebeling2010FinalMACS}; CLASH, \citealt{PostmanCLASHoverview}) and and SZ signatures (RELICS, \citealt{Coe2019RELICS,Acebron2019RXC0032,Paterno-Mahler2018SPT0615,Cerny2018}), or following optical signatures such as giant arcs \citep[e.g.,][]{Sharon2015RCS}. In contrast, and although RMJ1212 is ``only" modestly large -- note that it is larger than the typical Hubble Frontier Field cluster -- we stress that RMJ1212 was designated as a potentially large lens in a computerized, blind search in ground-based data following only the distribution and luminosity of cluster members as input (\citealt{Zitrin2012UniversalRE}, using the SDSS GMBCG cluster catalog of \citealt{Hao2010GMBCG_cat}). Here we confirm that, although somewhat smaller than the preliminary blind estimate of $\sim40\arcsec$, it is indeed a prominent lens. In Fig. \ref{fig:prelim} we show the preliminary critical curves predicted by the methodology and mass-to-light scaling of \citet{Zitrin2012UniversalRE}. We can compare these curves to the final curves presented here in Fig. \ref{fig:curve}, derived using the \emph{HST} data and careful multiple-image identification. Due to the overall successful assumption that light traces mass, the preliminary critical curves, derived from the SDSS data with no multiple images as input, are quite similar in shape to the final curves in Fig. \ref{fig:curve}, passing in between multiple images as they should and making RMJ1212 another proof-of-concept for identifying the largest lenses directly in ground-based and large sky surveys \citep{Zitrin2012UniversalRE,Wong2012OptLenses,Stapelberg2019EasyCritics}. This increasing ability to approximate the projected mass distribution and the corresponding lensing properties can be quite useful for upcoming surveys from the ground, or from space, such as with Euclid or the Nancy Grace Roman Space Telescope (previously known as WFIRST).

We can also compare the tentative multiple image identification in our ground-based Gemini data to the final identification presented here. \emph{Hubble} has the crucial combination of depth and high resolution -- a much better angular resolution than in typical ground-based observations (about $\sim0.1\arcsec$ compared to $\sim1\arcsec$) -- a key for securing the identification of multiple images, especially in lack of spectroscopic redshifts. Nevertheless, two systems and one candidate system that we initially identified in the GMOS data, guided by the location of the preliminary critical curves and following colors and symmetry, survived the more careful inspection allowed by the \emph{HST} data (at least partially, i.e., some counter images may have been updated). The other two system candidates we marked on the Gemini data seem to be wrong, emphasizing the need for space imaging for extensive multiple image identification. The HST data allowed us to detect, in addition, several other systems and multiple-image candidates, seen in Fig. \ref{fig:curve}.

\subsection{High-redshift candidates}\label{ss:highz}

We take advantage of the multiband observations and search the field for high-redshift candidates. Given that the field was observed for only about half an orbit in each WFC3/IR band, the target population are relatively, bright objects: the observing scheme was similar to that implemented in the RELICS program, designed to find bright ($5\sigma$ of AB 26.5 in the F160W band, for example) lensed candidates across $\sim$40 galaxy clusters. 

\begin{figure*}
 \begin{center}
 %\vspace{0.2cm}
  \textbf{ID25:}
  \includegraphics[width=188mm,trim=14cm 3cm 3cm 0cm,clip]{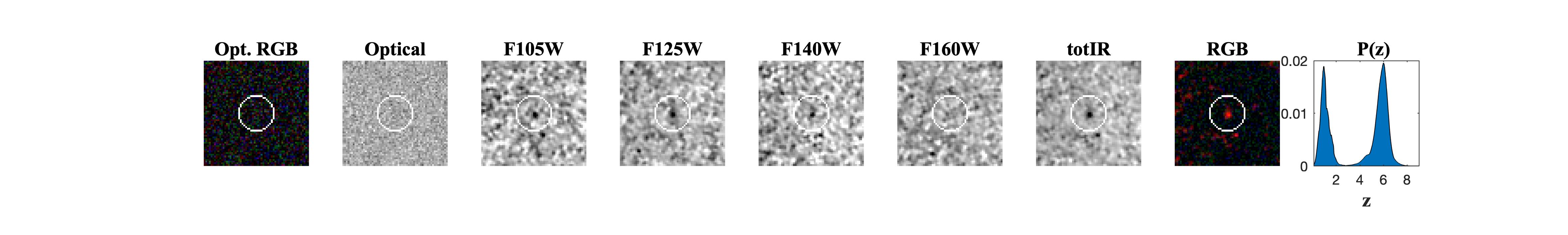}\\
   \textbf{ID244:}
   \includegraphics[width=188mm,trim=14cm 3cm 3cm 0cm,clip]{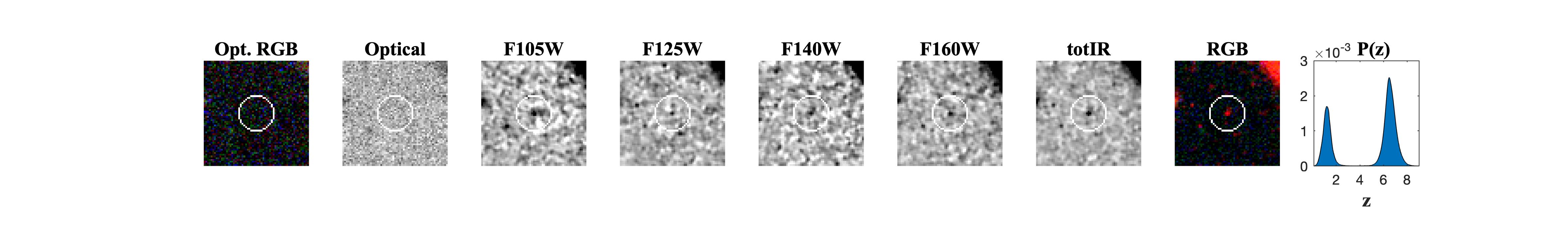}\\
    \textbf{ID311:}
    \includegraphics[width=188mm,trim=14cm 3cm 3cm 0cm,clip]{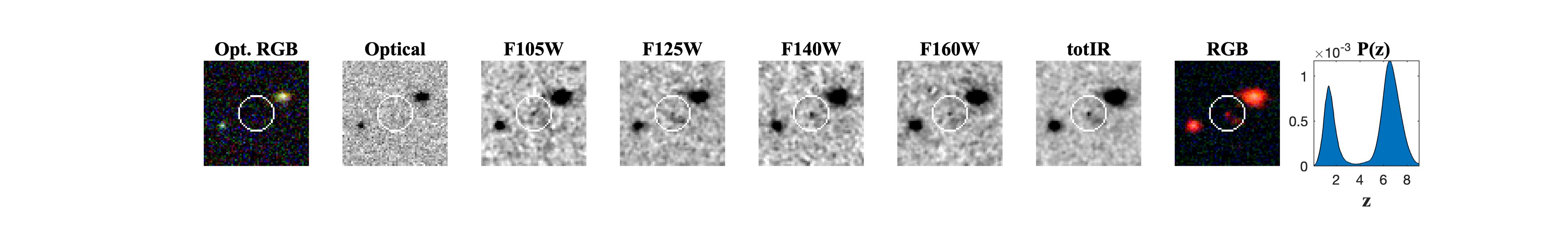}\\
     \textbf{ID613:}
        \includegraphics[width=188mm,trim=14cm 3cm 3cm 0cm,clip]{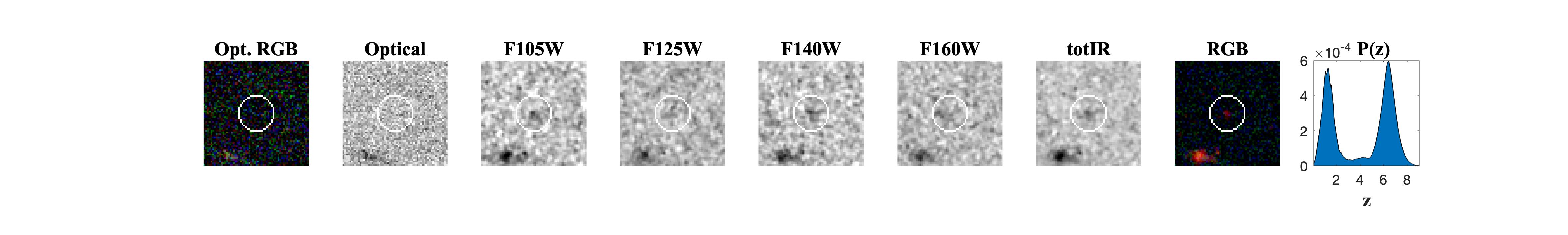}
      \textbf{ID544:}
     \includegraphics[width=188mm,trim=14cm 3cm 3cm 0cm,clip]{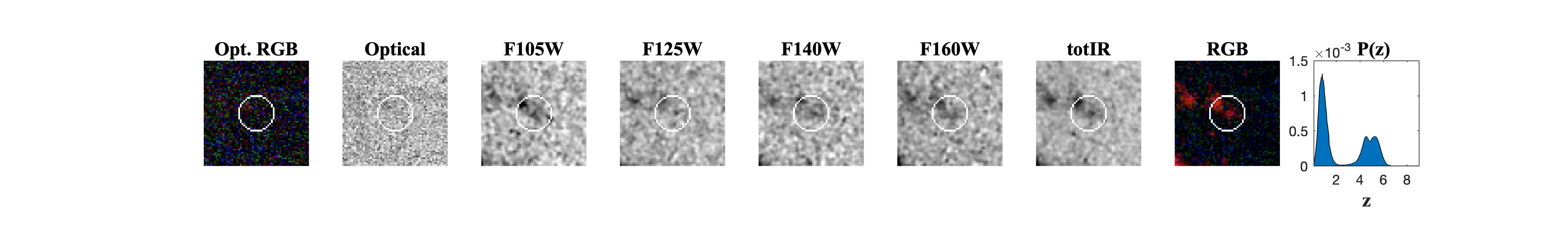}\\
  \textbf{ID1000:}
   \includegraphics[width=188mm,trim=14cm 3cm 3cm 0cm,clip]{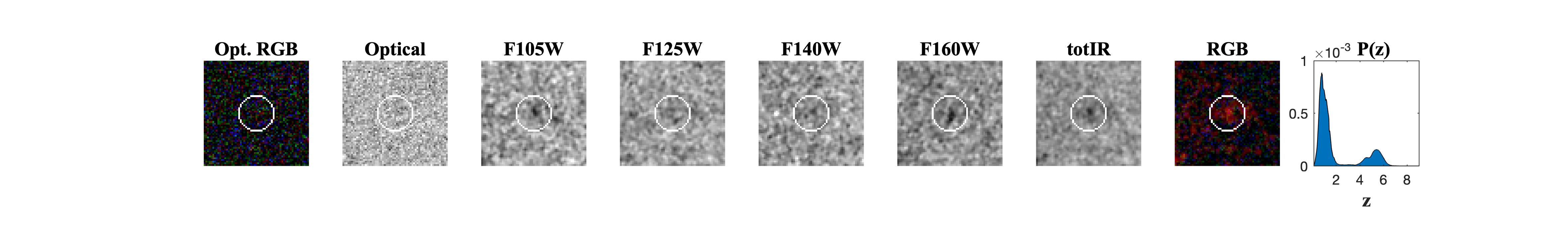}\\
   \textbf{ID1024:}
   \includegraphics[width=188mm,trim=14cm 3cm 3cm 0cm,clip]{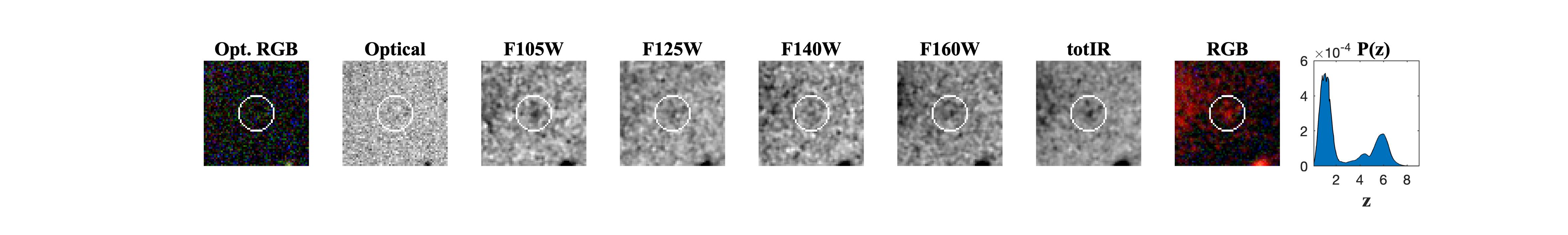}\\
 \end{center}
 \vspace{-0.3cm}
\caption{High-redshift candidates. Each row corresponds to a different object in the RELICS-like catalog, in the same order as in Table \ref{highztable}. For each object we show stamp images in the seven different bands, as well as in a combined optical (both in grey-scale and in a composite RGB image from the ACS bands), and combined RGB optical+infrared image. The first four objects are photo-$z$ selected and the rest of the objects were selected with the dropout technique and are undetected bluer of the Lyman break. Each stamp is $3.6\arcsec\times3.6\arcsec$ in size. Also shown is the photometric-redshift distribution for each object.}\vspace{0.1cm}
\label{fig:Dropout}
\end{figure*}

First, we search the RELICS-like BPZ catalog for objects with a $z_{best}>5.5$. Six objects pass this criterion. Two are designated as likely artifacts close ($<1\arcsec$) to the edge of the WFC3 frame and are discarded. The remaining four objects are listed in Table \ref{highztable}, and their stamp images in different bands are shown in Fig. \ref{fig:Dropout}. We then take on a second approach. We adopt the Lyman-break technique and apply selection criteria searching explicitly for dropout galaxies. Specifically, we adopt two sets of color criteria for the Lyman-break galaxy selection, as follows: \\
\\
The criteria used in \citet[][; hereafter criteria A]{Atek2014A2744}:\\
$\bullet$ Redshift $\sim6-7$ selection:\\
(F814W-F105W) $>$ 1.0\\
(F814W-F105W) $>$ (0.6 + 2.0*(F105W-F125W))\\
(F105W-F125W) $<$ 0.8\\
\\
$\bullet$ Redshift $\sim8$ selection:\\
(F105W-F125W) $>$ 0.5\\
(F105W-F125W) $>$ (0.3 + 1.6*(F125W-F140W))\\
(F125W-F140W) $<$ 0.5,\\
\\
and, the criteria used in \citet[][;hereafter criteria B]{Zheng2014A2744}:\\
$\bullet$ Redshift $\sim7-8$ selection:\\
(F814W - F105W) $>$ 0.8\\
(F814W - F105W) $>$ (0.8 + (F105W - F125W)) \\
(F105W - F125W) $<$ 0.6\\
\\
$\bullet$ Redshift $\sim8-9$ selection:\\
(F105W - F140W) $>$ 0.8\\
(F105W - F140W) $>$ (0.8 + (F140W - F160W))\\
(F140W - F160W) $<$ 0.6\\
\\
$\bullet$ Redshift $\sim10$ selection:\\
(F125W - F160W) $>$ 0.8,\\
\\
where F814W, F105W, F125W, F140W and F160W refer here to the magnitudes in those bands.
\\

We searched for objects that passed either selection, demanding in addition that these high-redshift dropout candidates are not detected by more than 2$\sigma$ in all bands bluewards of the break, as well as not detected by more than 1.5$\sigma$ in a weighted-stack image consisting of all bands bluer than the break. We only consider objects that are at least $1\arcsec$ away from the edge of the WFC3 frame, and adopt a \textsc{SExtractor} stellarity cut of $<0.8$ ($0=$galaxy, $1=$star), although we note that no object was removed due to this cut explicitly. We also discard objects that may be galactic brown-dwarf stars following the color-cut employed by \citep{Salmon2020RELICSHighz}. Finally, we require that any candidate be detected with at least 4$\sigma$ in the combined, WFC3/IR detection image.

\begin{figure*}
\begin{center}
      \textbf{ID117:}
     \includegraphics[width=188mm,trim=14cm 3cm 3cm 0cm,clip]{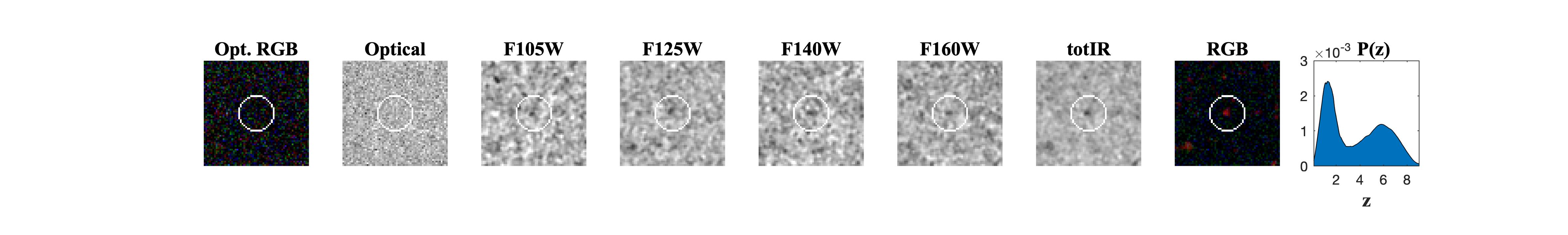}\\
   \textbf{ID304:}
   \includegraphics[width=188mm,trim=14cm 3cm 3cm 0cm,clip]{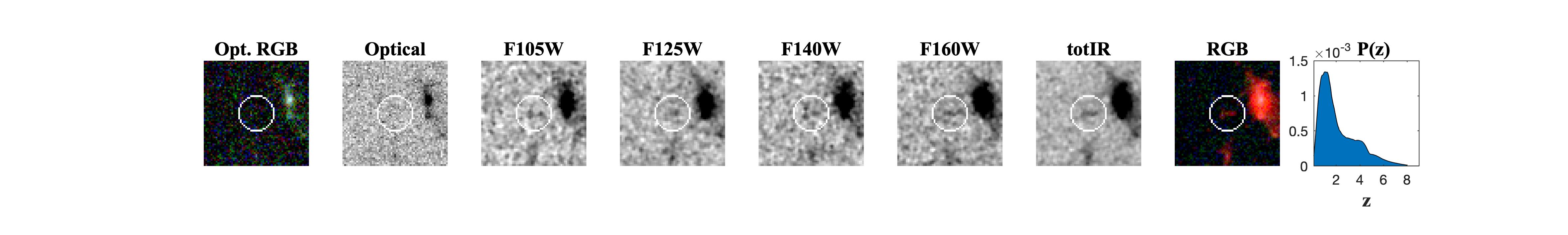}\\
     \textbf{ID732:}
   \includegraphics[width=188mm,trim=14cm 3cm 3cm 0cm,clip]{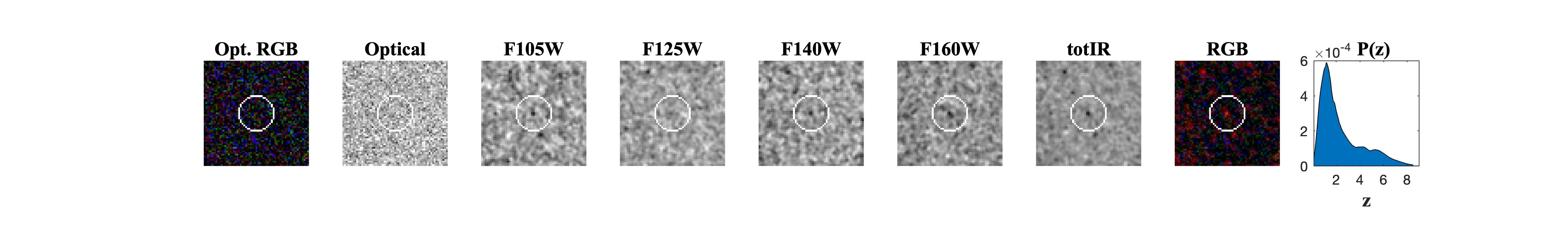}\\
   \textbf{ID821:}
     \includegraphics[width=188mm,trim=14cm 3cm 3cm 0cm,clip]{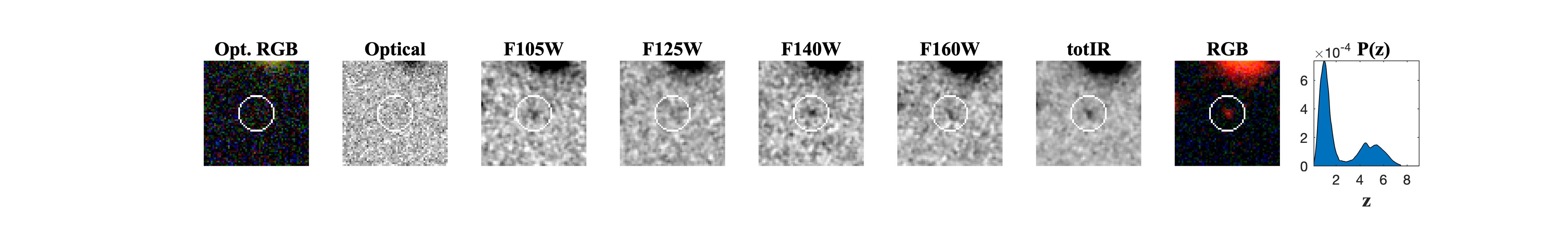}\\
  \textbf{ID897:}
     \includegraphics[width=188mm,trim=14cm 3cm 3cm 0cm,clip]{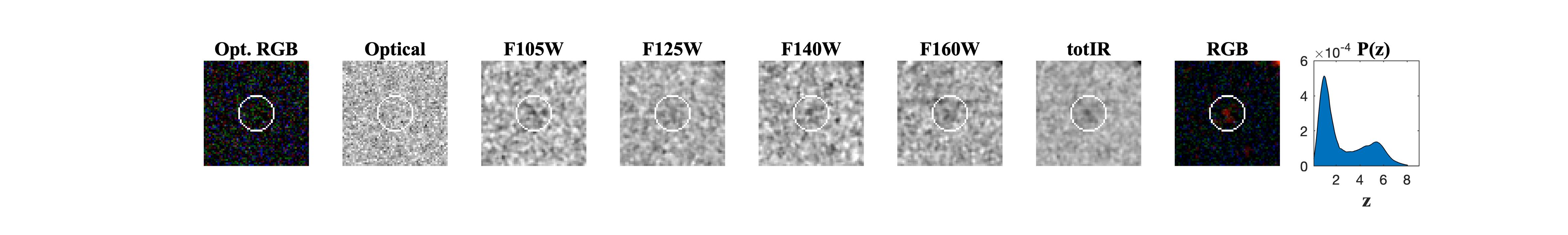}\\
  \textbf{ID958:}
   \includegraphics[width=188mm,trim=14cm 3cm 3cm 0cm,clip]{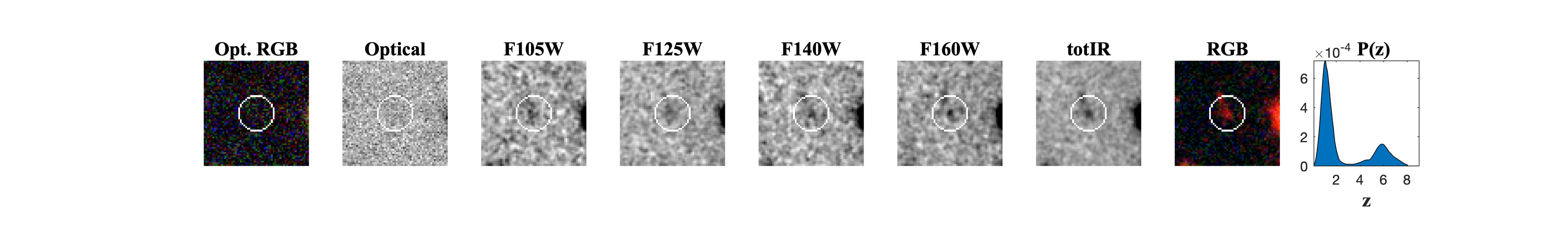}\\
   \textbf{ID998:}
     \includegraphics[width=188mm,trim=14cm 3cm 3cm 0cm,clip]{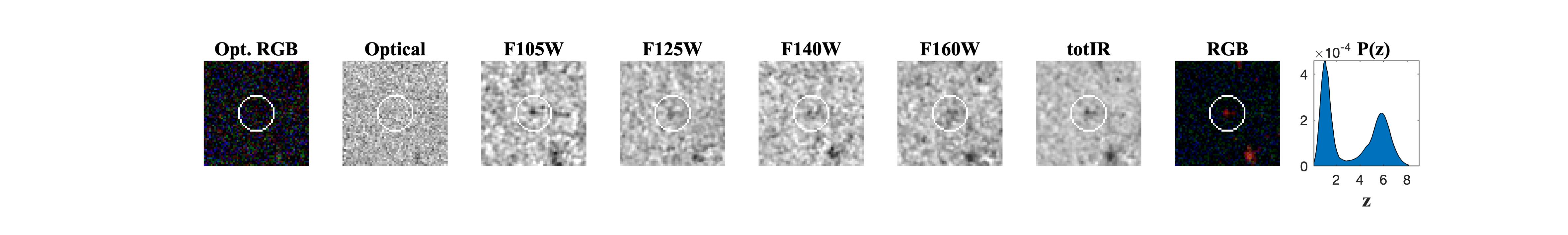}\\
  \end{center}
 \caption{Same as Fig. \ref{fig:Dropout}, but for dropout-selected candidates from our second catalog.}\vspace{0.1cm}
 \label{fig:Dropout2}
 \end{figure*}

We run these constraints by the RELICS-like ''wfc3ir" catalog. Three objects pass these criteria, and are listed in the second part of Table \ref{highztable}, and are shown in Fig. \ref{fig:Dropout}. We then also run these constraints by the alternative catalog we made for high-redshift source detection (\S \ref{s:data}; note that the lower $back\_size$ used here can lead to slightly different isophotal magnitudes compared to the RELICS-like catalog). Fourteen high-$z$ candidates pass these dropout selection criteria in this alternative catalog. Following a visual inspection by eye we discard three of them as likely artifacts (two are buried in the BCG light and one appears to be related to a lower-redshift counterpart). Four out of the remaining 11 overlap with candidates from the RELICS-like catalog (two from the photo-$z$ selected and two from the dropout-selected) so that overall, seven more objects are added to the list, listed in the third part of Table \ref{highztable} and shown in Fig. \ref{fig:Dropout2}, respectively. In total, we identify here 14 tentative high-redshift candidates. 

We note that, as seen in Figs. \ref{fig:Dropout}--\ref{fig:Dropout2}, the objects selected via the dropout technique all have a photometric-redshift distribution that allows for a high-redshift solution, but the best-fit suggests a lower redshift. In that sense it should be stressed that these candidates should be regarded with greater caution.

A few candidates are predicted by the model to be multiply imaged, such that others from the list may potentially be counter images of the same source. To search for counter images of the high-redshift candidates we lens each candidate image to the source-plane and back with the corresponding selection-criteria redshift, and check if other images from the list are within a radius of $3\arcsec$ from the predicted location of counter images. This practice indicates that objects ID8971 and ID311 may be related, but no other obvious counter images are found for other objects.

We can compare the high-$z$ yield to the high-redshift number counts in the RELICS survey, where the imaging scheme (filter choice and depth) is similar. \citet{Salmon2020HighzRelics} find 257, 57, and 8 candidate galaxies at $z_{phot}\simeq$ 6, 7, and 8, respectively, over 41 RELICS clusters, so that the average per cluster is 6.3, 1.4, and 0.2 galaxies at these redshifts, respectively, whereas the most prolific RELICS clusters can show above a couple-dozen candidates in total. In that sense, RMJ1212 seems to be comparable to, or somewhat stronger than the average RELICS cluster field in terms of high-$z$ number counts (although note the different selection methods). We emphasize that our list is preliminary, and our candidates, especially the fainter and smaller ones, will need more careful examination when more data is available. For example, the F160W (isophotal) magnitude distribution of the high-$z$ candidates of \citet{Salmon2020HighzRelics} concentrates around AB 27, with relatively few objects around AB 28. Our candidates seem to concentrate close to AB 28, comparable to the nominal depth limit ($\simeq1\sigma$ at AB $\simeq28.2$ for a point source). Note that we only intend to present here these high-z candidate galaxies to facilitate possible follow-up observations before the JWST mission. A more detailed examination of the high-redshift population behind this cluster is warranted, and remains for future work.

\subsection{Transients}
Last, we also take advantage of the fact that the cluster was imaged in two different epochs to search for transient sources such as potential supernovae or caustic crossing events. The WFC3-IR integrations through the F105W, F125W, F140W, and F160W
wideband filters were acquired first on March 16 2020, UT and at second epoch, nine days later, on March 25, 2020 UT (the ACS WFC imaging was acquired, by contrast, at a single epoch). We have searched the two epochs of WFC3 IR imaging data for variable sources, including supernovae and microlensing events caused as the caustic magnification pattern in the source plane moves relative to the stars in a magnified arc. While microlensing due to a moving caustic pattern will, in general, cause a continuous change in magnification, the characteristic time scale of microlensing {\it peaks} \citep[e.g.,][]{Kelly2018NatAsCCE,Rodney2018NatAsCCE,Chen2019CCE,Kaurov2019CCE} should be approximately two weeks, which was confirmed by that of the Icarus event in MACS\,J1149 \citep{Kelly2018NatAsCCE}.  According to the WFC3 Exposure Time Calculator (ETC), the 5-$\sigma$ AB detection limits are approximately 26.0 mag (F105W), 25.9 (F125W), 25.9 (F140W), and 26.0 (F160W), after taking into account the background-noise from the template image. For added sensitivity, we also carried out a search of the coaddition of all WFC3 filters. Visual inspection and SExtractor searches, however, yield no credible ($\gtrsim5\sigma$) transients either at the locations of the prominent arcs or at other locations in the imaging footprint. 
Nevertheless, as the cluster shows several arcs that clearly straddle the critical curves, such as the blue star-forming systems 1-3, or systems 4 and c5, it should in principle be useful for future caustic crossing searches.

\section{Summary}\label{s:summary}

We presented a SL model for the very rich redMaPPER galaxy cluster RMJ1212 (also known as Abell 1489,
RXC J1212.3+2733, or CL1212+2733), in preparation for the WMDF JWST/GTO program (\# 1176) that is planned to observed this cluster with NIRCam on JWST. In recent Hubble multi-band ACS+WFC3/IR imaging we have identified 7 sets of multiple-image sets that were used as constraints, as well as several less secure candidates, and reveal a prominent lens of $\theta_{E}\simeq32\pm3\arcsec$ ($z_{s}=2$), and $\theta_{E}\simeq39\pm4\arcsec$ ($z_{s}=10$). 

We searched the data for high-redshifts candidates. We found four candidate high-redshift objects ($z\gtrsim6$) based on a photometric-redshift selection. Applying independently a dropout, Lyman break selection criteria, we uncovered 10 more tentative objects. While we note that our candidates require an additional, future examination, especially when more data become available, these numbers are typical of lensing clusters imaged to similar depths \citep{Salmon2020HighzRelics}. We also searched the data for transient sources. No significant transients were found between the two HST visits (separated by nine days). Nevertheless, there are a few arcs that cross the critical curve and should be useful for caustic crossing searches in future imaging of this cluster, especially with JWST.

RMJ1212 was provisionally chosen for follow-up based on a relatively large Einstein radius implied from mass-to-light scaling relations in SDSS data. Our analysis here reveals a somewhat ($\sim20\%$) smaller lens than predicted by the scaling relations, but overall the size and shape of the critical curves are in broad agreement with these blind predictions. Only about half of the multiple image candidates that we had identified in ground-based data survived the detailed analysis presented here, emphasizing the need for \emph{Hubble} data for lens modeling and multiple-image identification. In addition, the lensing analysis has revealed that most of the mass is concentrated around the second and third central BCGs, whereas the brightest cluster member seems to be a much smaller concentration of mass than implied by its luminosity. The overall success of our automated procedures to flag RMJ1212 and approximate its lensing properties based solely on the photometry and distribution of cluster members in SDSS data, is another example of a growing ability to map large numbers of cluster lenses automatically in large sky surveys \citep{Zitrin2012UniversalRE}. It is not hard to imagine that the combination of such methods \citep{Carrasco2020,Stapelberg2019EasyCritics,Wong2012OptLenses} including increasingly-popular machine learning techniques, with wide-field space data as expected from Euclid or the Roman Space Telescope, will enable in a few years time fully automated and increasingly robust lensing analyses of large samples of clusters.

\section*{acknowledgements}
M.N. acknowledges INAF 1.05.01.86.20 and PRIN MIUR F.OB. 1.05.01.83.08. J.M.D. acknowledges the support of project PGC2018-101814-B-100 (MCIU/AEI/MINECO/FEDER, UE) Ministerio de Ciencia, Investigaci\'on y Universidades. This project was funded by the Agencia Estatal de Investigaci\'on, Unidad de Excelencia Mar\'ia de Maeztu, ref. MDM-2017-0765. 

This work is based on observations made with the NASA/ESA Hubble Space Telescope obtained from the Space Telescope Science Institute, which is operated by the Association of Universities for Research in Astronomy, Inc., under NASA contract NAS 5–26555. These observations are associated with program ID 15959.

RAW, SHC and RAJ acknowledge support from NASA JWST Interdisciplinary Scientist
grants NAG5-12460, NNX14AN10G and 80NSSC18K0200 from GSFC.

Some scientific results reported in this article are based in part on observations made by the Chandra X-ray Observatory.
This work is also based in part on observations obtained at the international Gemini Observatory, a program of NSF’s OIR Lab, which is managed by the Association of Universities for Research in Astronomy (AURA) under a cooperative agreement with the National Science Foundation on behalf of the Gemini Observatory partnership: the National Science Foundation (United States), National Research Council (Canada), Agencia Nacional de Investigaci\'{o}n y Desarrollo (Chile), Ministerio de Ciencia, Tecnolog\'{i}a e Innovaci\'{o}n (Argentina), Minist\'{e}rio da Ci\^{e}ncia, Tecnologia, Inova\c{c}\~{o}es e Comunica\c{c}\~{o}es (Brazil), and Korea Astronomy and Space Science Institute (Republic of Korea). These guaranteed observations were obtained through Ben-Gurion University's (BGU; Israel) time on Gemini, following a MoU between BGU and Gemini/AURA.

Funding for the Sloan Digital Sky Survey IV has been provided by the Alfred P. Sloan Foundation, the U.S. Department of Energy Office of Science, and the Participating Institutions. SDSS-IV acknowledges support and resources from the Center for High-Performance Computing at the University of Utah. The SDSS web site is www.sdss.org.

SDSS-IV is managed by the Astrophysical Research Consortium for the Participating Institutions of the SDSS Collaboration including the  Brazilian Participation Group, the Carnegie Institution for Science,  Carnegie Mellon University, the Chilean Participation Group, the French Participation Group, Harvard-Smithsonian Center for Astrophysics,  Instituto de Astrof\'isica de Canarias, The Johns Hopkins University, Kavli Institute for the Physics and Mathematics of the Universe (IPMU) / 
University of Tokyo, the Korean Participation Group, Lawrence Berkeley National Laboratory, 
Leibniz Institut f\"ur Astrophysik Potsdam (AIP),  
Max-Planck-Institut f\"ur Astronomie (MPIA Heidelberg), 
Max-Planck-Institut f\"ur Astrophysik (MPA Garching), 
Max-Planck-Institut f\"ur Extraterrestrische Physik (MPE), 
National Astronomical Observatories of China, New Mexico State University, 
New York University, University of Notre Dame, 
Observat\'ario Nacional / MCTI, The Ohio State University, 
Pennsylvania State University, Shanghai Astronomical Observatory, 
United Kingdom Participation Group,
Universidad Nacional Aut\'onoma de M\'exico, University of Arizona, 
University of Colorado Boulder, University of Oxford, University of Portsmouth, 
University of Utah, University of Virginia, University of Washington, University of Wisconsin, 
Vanderbilt University, and Yale University.

%This work is based on observations made with the NASA/ESA Hubble Space Telescope. Support for key programs that enabled this work (mainly \#12065, 13504, 134509, 14041, 13790, 14199) was provided by NASA from the Space Telescope Science Institute (STScI), which is operated by the Association of Universities for Research in Astronomy (AURA), Inc. under NASA contract NAS 5-26555.

\newpage
\movetabledown=6.5in
%\movetableright=1in
\begin{rotatetable*}
\begin{deluxetable}{lcccccccccccccc}
\tablecaption{High-Redshift Candidates}
\label{highztable}
%\tablecolumns{13}
\footnotesize
\tabletypesize{}
\tablewidth{1\linewidth}
\tablehead{
\colhead{ID
} &
\colhead{R.A
} &
\colhead{DEC.
} &
\colhead{$mag_{F435W}$
} &
\colhead{$mag_{F606W}$
} &
\colhead{$mag_{F814W}$
} &
\colhead{$mag_{F105W}$
} &
\colhead{$mag_{F125W}$
} &
\colhead{$mag_{F140W}$
} &
\colhead{$mag_{F160W}$
} &
\colhead{$z_{phot}$ [95\% C.I.]
} &
\colhead{$\mu$}\\  
 &J2000.0&J2000.0&&&&& 
}
\startdata
\hline
ID25  & 12:12:20.259 &  +27:34:28.37 &--& $ 29.85 \pm 0.69$ & $ 28.76\pm 0.44$ & $ 27.25\pm 0.26$& $ 26.96 \pm 0.27$& $27.37 \pm 0.31$ & $ 28.71\pm 0.82$ & 6.01 [0.50 -- 6.70] &  3.90 $\pm$ 0.05\\
ID244 & 12:12:15.230 & +27:33:43.22& -- & -- & --& $ 27.45\pm 0.26$ & $28.11 \pm0.52 $& $27.84 \pm 0.37$& $ 27.73 \pm 0.33$ & 6.48 [0.73 -- 7.50]& 6.53 $\pm$ 0.13 \\
ID311$^{a67,b78}$ & 12:12:21.822 & +27:33:36.59& $ 29.54 \pm 0.72$& -- &$ 32.72 \pm 3.06$ & $ 28.22\pm 0.39$ & $ 28.17\pm 0.47$& $ 27.92\pm 0.34$& $28.35 \pm0.44 $& 6.52 [ 0.80 -- 8.18]& 2.58 $\pm$ 0.03\\
ID613$^{a67,b78}$ & 12:12:23.615 & +27:33:11.33 & -- & $30.29\pm 0.96$&-- & $ 27.95 \pm 0.31$ &$ 28.37 \pm 0.52$ & $ 27.76\pm 0.29$& $28.88 \pm 0.61$ & 6.43 [0.56 -- 7.58]&  2.12 $\pm$ 0.02\\
\hline
\hline
ID544$^{b78}$ & 12:12:20.861 & +27:33:15.60 & --- & $29.10 \pm 0.57$ & $27.83 \pm 0.30$ & $26.97 \pm 0.21$ & $27.70 \pm 0.47$ & $27.16 \pm 0.27$ & $27.16 \pm 0.26$ & 0.75 [0.37 -- 5.75] & 4.42 $\pm$ 0.08 \\
ID1000$^{b78}$ & 12:12:17.112 & +27:32:34.75 & --- & $29.14 \pm 0.68$ & $27.84 \pm 0.36$ & $26.89 \pm 0.23$ & $26.98 \pm 0.31$ & $27.29 \pm 0.34$ & $26.59 \pm 0.18$ & 0.83 [0.41 -- 5.91] & 10.13 $\pm$ 0.31 \\
ID1024$^{a67,b78}$ & 12:12:17.258 & +27:32:33.06 & --- & $29.19 \pm 0.59$ & $28.79 \pm 0.62$ & $27.19 \pm 0.25$ & $27.22 \pm 0.32$ & $27.81 \pm 0.43$ & $27.45 \pm 0.31$ & 1.06 [0.40 -- 6.54] & 13.29 $\pm$ 0.52\\
\hline
\hline
% ID81 & 12:12:20.1648 & +27:32:04.081 & --- & $30.97 \pm 2.11$ & $32.84 \pm 17.29$ & $28.82 \pm 0.62$ & $28.91 \pm 0.88$ & $27.74 \pm 0.25$ & $28.40 \pm 0.44$ & 1.42 [0.51 -- 8.68] & 5.00 $\pm$ 0.09\\
ID117$^{b78}$ & 12:12:19.426 & +27:32:10.42 & $29.75 \pm 1.21$ & --- & $29.88 \pm 1.25$ & $28.52 \pm 0.50$ & $28.10 \pm 0.44$ & $28.14 \pm 0.38$ & $28.37 \pm 0.45$ & 1.28 [0.47 -- 8.01] & 10.49 $\pm$ 0.36\\
%ID278(67 1,2) & 12:12:17.258 & +27:32:33.07 & --- & $29.44 \pm 0.71$ & $29.12 \pm 0.79$ & $27.65 \pm 0.28$ & $27.65 \pm 0.37$ & $28.20 \pm 0.52$ & $27.89 \pm 0.37$ & 1.05 [0.33 -- 6.56] & 13.29 $\pm$ 0.52\\
ID304$^{a67,b78}$ & 12:12:21.763 & +27:32:36.55 & $31.03 \pm 5.56$ & $29.10 \pm 0.55$ & $29.16 \pm 0.97$ & $28.03 \pm 0.44$ & $27.81 \pm 0.47$ & $28.16 \pm 0.55$ & $27.75 \pm 0.36$ & 1.03 [0.30 -- 6.09] & 6.63 $\pm$ 0.18\\
% ID706(67 1,2) &12:12:23.616 & +27:33:11.33 & --- & $29.99 \pm 1.26$ & --- & $27.83 \pm 0.32$ & $28.27 \pm 0.64$ & $27.59 \pm 0.29$ & $28.68 \pm 0.74$ & 1.09 [0.54 -- 7.53] &2.13 $\pm$ 0.02\\
ID732$^{a67,b78}$ & 12:12:16.668 & +27:33:13.40 & --- & $30.02 \pm 0.87$ & $29.62 \pm 0.88$ & $28.55 \pm 0.48$ & $28.80 \pm 0.80$ & $28.35 \pm 0.44$ & $28.18 \pm 0.36$ & 1.17 [0.36 -- 6.79] & 160.47 $\pm$ 1993.08$^{*}$\\
%ID761(67,2) & 12:12:20.861 & +27:33:15.60 & --- & $29.25 \pm 0.80$ & $27.97 \pm 0.37$ & $27.05 \pm 0.23$ & $27.70 \pm 0.54$ & $27.22 \pm 0.29$ & $27.26 \pm 0.29$ & 0.81 [0.35 -- 5.98] & 4.42 $\pm$ 0.08\\
ID821$^{a67,b78}$ & 12:12:14.815 & +27:33:20.00 & --- & $29.16 \pm 0.61$ & $28.73 \pm 0.81$ & $27.35 \pm 0.25$ & $28.01 \pm 0.61$ & $27.18 \pm 0.24$ & $27.58 \pm 0.32$ & 0.98 [0.37 -- 6.46] & 4.45 $\pm$ 0.07\\
ID897$^{b78}$ & 12:12:16.176 & +27:33:26.26 & --- & $31.10 \pm 3.11$ & $29.11 \pm 0.65$ & $28.14 \pm 0.38$ & $28.45 \pm 0.68$ & $28.05 \pm 0.39$ & $28.31 \pm 0.47$ & 0.98 [0.34 -- 6.63] & 46.35 $\pm$ 5.49\\
ID958$^{a67,b78}$ & 12:12:19.550 & +27:33:33.93 & --- & $29.40 \pm 0.88$ & $28.74 \pm 0.71$ & $27.43 \pm 0.31$ & $27.14 \pm 0.31$ & $27.02 \pm 0.23$ & $27.20 \pm 0.26$ & 1.06 [0.47 -- 7.09] & 156.03 $\pm$ 2353.34$^{*}$\\
% ID993(67, 1,2) & 12:12:21.823 & +27:33:36.60 & $29.89 \pm 1.30$ & --- & $32.79 \pm 16.86$ & $28.26 \pm 0.43$ & $28.25 \pm 0.56$ & $28.10 \pm 0.41$ & $28.40 \pm 0.51$ & 1.27 [0.46 -- 7.80] & 2.58 $\pm$ 0.03\\
ID998$^{a67,b78}$ & 12:12:21.240 & +27:33:37.01 & --- & $30.25 \pm 1.54$ & $29.24 \pm 0.90$ & $27.63 \pm 0.30$ & $27.65 \pm 0.40$ & $27.74 \pm 0.37$ & $27.80 \pm 0.37$ & 1.04 [0.46 -- 7.10] & 3.00 $\pm$ 0.04\\
\hline
\enddata
\tablecomments{High-redshift ($z\gtrsim6$) galaxy candidates. \emph{Column~1:} ID; \emph{Column~2 \& 3:} Right Ascension and Declination, in J2000.0; \emph{Column~4 - 10:} isophotal magnitudes and associated uncertainty measured by Source-EXtractor; \emph{Column~11:} best photometric redshift from BPZ, and its 95\% confidence interval; \emph{Column~12:} approximate magnification by the model, adopting the relevant photometric or dropout-selection redshift. $^{*}$ Diverging values suggest the object is close to the critical curves, highly
magnified but with poorly constrained magnification.\\
The first part of the Table are objects selected by considering entries with $z_{phot}>5.5$ in the automated RELICS-like catalog. The second part of the Table are objects that passed the dropout-selection criteria in the RELICS-like catalog (with no photometric redshift cut; see \S \ref{ss:highz}), and the third part are objects that passed the dropout-selection criteria in our alternative, designated catalog. Note this catalog used different SExtractor parameters so the isophotal magnitudes can be slightly (typically $\sim 0.1-0.2$ mag) different compared to the RELICS-like catalog. For each galaxy we note which criteria it passed (''a"=A or ''b"=B), and for which redshift (''67" stands for $z\sim6-7$; ''78" for $z\sim7-8$, ''8" for $z\sim8$, etc.). Two of the $photo-z$ selected galaxies were also recovered by the dropout selection. See text for more details.}% \emph{Column~11:} comments.}
\end{deluxetable}
\end{rotatetable*}

\newpage

%\bibliographystyle{aasjournal}
%\bibliography{bibfile}

\end{document}